\documentclass[aps, twocolumn, prd, superscriptaddress, 
showpacs,10pt]{revtex4-1}

\usepackage{amsmath} \usepackage{amsfonts} \usepackage{amssymb}
\usepackage{amsthm} \usepackage{mathtools} \usepackage{mathrsfs}
\usepackage{txfonts} 
\usepackage{multirow}
\usepackage{color} \usepackage{xspace}
\usepackage[caption=false,singlelinecheck=false]{subfig} \usepackage{epsfig}
\usepackage{graphicx}

\usepackage{tikz} \usetikzlibrary{calc} \usetikzlibrary{decorations}
\usetikzlibrary{decorations.text} \usetikzlibrary{decorations.fractals}
\usetikzlibrary{decorations.pathreplacing}
\usetikzlibrary{decorations.pathmorphing} \usetikzlibrary{decorations.markings}
\usetikzlibrary{shadings} \usetikzlibrary{positioning,matrix}
\usetikzlibrary{shapes.arrows} \usetikzlibrary{shapes.symbols}
\usetikzlibrary{shapes.misc} \usetikzlibrary{shapes.geometric}
\usetikzlibrary{through} \usetikzlibrary{mindmap,backgrounds}
\usetikzlibrary{fit} \usetikzlibrary{arrows,positioning}
\usetikzlibrary{external} \usetikzlibrary{spy}

\usepackage[colorlinks, urlcolor=black, citecolor=blue, link color=black]{hyperref}

\def \Re{\text{Re}}

\def\Dbar {\kern 0.2em\bar{\kern -0.2em D}{}\xspace}

\def\Dzb   {\ensuremath{\Dbar^0}\xspace}
\def\DzDzb {\ensuremath{D^0 {\kern -0.16em \Dzb}}\xspace}
\def\Dp    {\ensuremath{D^+}\xspace}

\def\Dzzb {\kern 0.2em\overset{(\bar{\kern -0.2em D})}{}\xspace}

\def\Dzbs   {\ensuremath{\Dbar^{*0}}\xspace}
\def\DzsDzbs {\ensuremath{D^{*0} {\kern -0.16em \Dzbs}}\xspace}

\def\Bbar  {\kern 0.18em\bar{\kern -0.18em B}{}\xspace}

\def\Bzb   {\ensuremath{\Bbar^0}\xspace}
\def\BzBzb {\ensuremath{B^0 {\kern -0.16em -\Bzb}}\xspace}
\def\Bp    {\ensuremath{B^+}\xspace}

\def\Bz    {\ensuremath{B^0}\xspace}
\def\CP {\ensuremath{C\!P}\xspace}
\def\CPT {\ensuremath{C\!P T}\xspace}

\def\Kbar  {\kern 0.2em\bar{\kern -0.2em K}{}\xspace}

\def\Kzb   {\ensuremath{\Kbar^0}\xspace}

\def\Kp    {\ensuremath{K^+}\xspace}
\def\Km    {\ensuremath{K^-}\xspace}

\def\Kz    {\ensuremath{K^0}\xspace}

\def\Bbar  {\kern 0.18em\bar{\kern -0.18em B}{}\xspace}

\def\Bzb   {\ensuremath{\Bbar^0}\xspace}
\def\BzBzb {\ensuremath{B^0 {\kern -0.16em -\Bzb}}\xspace}

\def\nn    {\nonumber}
\def\sss{\scriptscriptstyle}

\def\SU{\ensuremath{S\!U}}

\def\pip    {\ensuremath{\pi^+}\xspace}
\def\pim    {\ensuremath{\pi^-}\xspace}

\def\piz    {\ensuremath{\pi^0}\xspace}

\def\e{{\rm e}}
\def\o{{\rm o}}

\def\half	{\ensuremath{\frac{1}{2}}}
\def\thalf	{\ensuremath{\frac{3}{2}}}
\newcommand{\T}[1]	{\ensuremath{T_{\!\sss{1,\frac{#1}{2}}}}}
\def\t		{\ensuremath{T_{\!\sss{0,\frac{1}{2}}}}}
\def\u		{\ensuremath{U_{\!\sss{\frac{1}{2},\frac{1}{2}}}}}
\newcommand\up[1]		{\ensuremath{U^{\prime\,#1}_{\!\sss{\frac{1}{2},\frac{1}{2}}}}}
\def\v		{\ensuremath{V_{\!\sss{\frac{1}{2},\frac{1}{2}}}}}
\newcommand\vp[1]		{\ensuremath{V^{\prime\,#1}_{\!\sss{\frac{1}{2},\frac{1}{2}}}}}

\def \GI {\ensuremath{G_{\! I}}}
\def \GU {\ensuremath{G_{\! U}}}
\def \GV {\ensuremath{G_{\! V}}}

\newcommand{\bracket}[3]{\left\langle #1 \right\lvert #2 \left\rvert #3 
\right\rangle}

\newcommand{\biggbracket}[3]{\bigg\langle #1\bigg\lvert #2\bigg\rvert
#3\bigg\rangle}
\newcommand{\ket}[1]{\left| #1 \right\rangle}

\newcommand{\modulus}[1]{\left| #1 \right|}

\def\sss{\scriptscriptstyle}
\def\dsp{\displaystyle}

\def\bBar#1{\hbox{$#1$\kern -1.85em\raise1.6ex\hbox{{\raise.35ex
\hbox{$~{\sss(}$}}$-${\raise.35ex\hbox{${\sss )}$}}}\kern 0.3em}}

\makeatletter
\def\equalsfill{$\m@th\mathord=\mkern-7mu
\cleaders\hbox{$\!\mathord=\!$}\hfill
\mkern-5mu\mathord=$}
\makeatother

\newcommand{\eqst}{\stackrel{\ensuremath{s \leftrightarrow t}}{\hbox{\equalsfill}}}%
\newcommand{\eqtu}{\stackrel{\ensuremath{t \leftrightarrow u}}{\hbox{\equalsfill}}}%

\newcommand{\ASS}{\ensuremath{\mathcal{A}_{SS}}}
\newcommand{\ASA}{\ensuremath{\mathcal{A}_{SA}}}
\newcommand{\AAS}{\ensuremath{\mathcal{A}_{AS}}}
\newcommand{\AAA}{\ensuremath{\mathcal{A}_{AA}}}

\allowdisplaybreaks

\begin{document}

\title{A model independent method for quantitative estimation of
\texorpdfstring{$\SU(3)$}{SU(3)} flavor symmetry  breaking using Dalitz plot}

\author{Dibyakrupa Sahoo} 
\affiliation{The Institute of Mathematical Sciences, Taramani,
Chennai 600113, India}
\author{Rahul Sinha} 
\affiliation{The Institute of Mathematical Sciences, Taramani,
Chennai 600113, India}
\author{N.~G.~Deshpande}
\affiliation{Institute of Theoretical Science, University of Oregon,
Eugene, Oregon 94703, USA}

\date{\today}

\begin{abstract}
The light hadron states are satisfactorily described in the quark model using
$\SU(3)$ flavor symmetry. If the $\SU(3)$ flavor symmetry relating the light
hadrons were exact, one would have an exchange symmetry between these hadrons
arising out of the exchange of the up, down and strange quarks. This aspect of
$\SU(3)$ symmetry is  used extensively to relate many decay modes of heavy
quarks. However, the nature of the effects of $\SU(3)$ breaking in such decays
is not well understood and hence, a reliable estimate of $\SU(3)$ breaking
effects is missing. In this work we propose a new method to quantitatively
estimate the extent of flavor symmetry breaking and better understand the nature
of such breaking using Dalitz plot. We study the three non-commuting $\SU(2)$
symmetries (subsumed in $\SU(3)$ flavor symmetry): isospin (or $T$-spin),
$U$-spin and $V$-spin, using the Dalitz plots of some three-body meson decays.
We look at the Dalitz plot distributions of decays in which pairs of the final
three particles are related by two distinct $\SU(2)$ symmetries. We show
that such decay modes have characteristic distributions that enable the
measurement of violation of each of the three $\SU(2)$ symmetries via Dalitz
plot asymmetries in a single decay mode.  Experimental estimates of these
easily measurable asymmetries  would help in better understanding the weak
decays of heavy mesons into both two and three light mesons.
\end{abstract}

\maketitle

\section{Introduction}

A satisfactory understanding of the light hadronic states using $\SU(3)$ flavor
symmetry is one of the outstanding success stories of particle
physics~\cite{GellMann:1961, GellMann:1962xb, Ne'eman:1961cd, Okubo:1961jc,
Okubo:1964xz}. In its true essence the $\SU(3)$ flavor symmetry denotes the full
exchange symmetry amongst the up ($u$), down ($d$) and strange ($s$) quarks.
Another implication of $\SU(3)$ flavor symmetry, if it were an exact symmetry,
is that the mesons formed by combining the quarks $u$, $d$, $s$ and the
antiquarks $\bar u$, $\bar d$, $\bar s$ belonging to the same representation of
$\SU(3)$ would also be degenerate. One treats the three quarks on the same
footing even though the quark masses differ by allowing for a breaking of the
symmetry. The success of the Gell-Mann-Okubo mass formula in relating the hadron
masses is that it takes the small $\SU(3)$ breaking into account but does not
depend on the details of $\SU(3)$ breaking effects. Such $\SU(3)$ breaking
effects cannot be calculated and must be estimated using experimental inputs.
Traditionally, the mass differences between these mesons have been used as a
measure of the extent of breaking of $\SU(3)$ flavor symmetry. The masses of
these mesons, which are bound states of quark-antiquark pairs, depend on their
binding energies. It is not possible to estimate these binding energies from QCD
calculations since these resonances lie in the non-relativistic low energy
regime. Moreover, the electro-magnetic interactions between the quark and the
antiquark in the meson also contribute towards its binding energy. Thus, by
measuring the mass differences amongst the mesons one does not fully solicit the
breaking of $\SU(3)$ flavor symmetry. Another usual way to explore the breaking
$\SU(3)$ flavor symmetry is to look at specific loop diagrams where the down and
strange quarks contribute. The loop effects affect the amplitude of the process
under consideration and its physical manifestations are then studied for a
quantitative estimation of the breaking of $\SU(3)$ flavor symmetry. Since up
quark has different electric charge than down and strange, it can not be treated
in the same way in these studies of loop contributions. Therefore, such a method
also fails to probe the full exchange symmetry of these three light quarks.
Hence, all estimates of $\SU(3)$ breaking are currently empirical.

Several studies exist in literature that have used broken $\SU(3)$ flavor
symmetry (i) in various decay modes using the methods of amplitudes (usually
isospin and $U$-spin amplitudes) and various quark diagrams
~\cite{Kingsley:1975fe, Voloshin:1975yx, Wang:1979dx, Quigg:1979ic,
Zeppenfeld:1980ex, Chau:1986du, Chau:1987tk, Savage:1989ub, Chau:1990ay,
Savage:1991wu, Lipkin:1991st, Chau:1991gx, Kwong:1993ri, Hinchliffe:1995hz,
Gronau:1995hm, Oh:1998wa, Rosner:1999xd, Gronau:2000ru, Deshpande:2000jp,
Wu:2002nz, Khodjamirian:2003xk, Chiang:2003pm, Zhong:2004ck, Chiang:2004nm,
Wu:2004ht, Wu:2005hi, Gronau:2005ax, Chiang:2006ih, Soni:2006vi, Chiang:2007bd,
Chiang:2007qh, Bhattacharya:2008ss, Chiang:2008zb, Wei:2009zzh, Jung:2009pb,
Imbeault:2011jz, Cheng:2011qh, Pirtskhalava:2011va, Cheng:2012xb, Pham:2012db,
Hiller:2012xm, Bhattacharya:2014eca, Cheng:2014rfa, Pham:2014lla, He:2014xha,
Ledwig:2014rfa, Gronau:2015rda}, and (ii) in determinations of weak phases and
$\CP$ violating phases~\cite{Kim:1998sh, Gronau:1998rr, Datta:2003va,
Gronau:2005pq, Gronau:2006gn, Gronau:2007af, Calibbi:2009pv, ReyLeLorier:2011ww,
He:2013vta, Bhattacharya:2013cla, Grossman:2013lya, Fong:2013dnk}. These methods
involve comparison of observables in distinct decay modes which are related by
some underlying $\SU(2)$ symmetries, such as isospin, $U$-spin or $V$-spin.
However, the full exchange symmetry amongst the three light quarks has not yet
been fully exploited, in a single decay mode. Hadronic weak decays involve
several unknown parameters which can be reduced by the use of $\SU(3)$ flavor
symmetry. Since, $\SU(3)$ flavor symmetry  is still extensively used to relate
the few decay modes of heavy quarks, it is important to realize  other ways to
experimentally measure the breaking of $\SU(3)$ flavor symmetry and understand
better the complete nature of $\SU(3)$ breaking. In this paper we propose a
method to achieve precisely this by looking at asymmetries in the Dalitz plot
under exchange of the mesons in the final state. These asymmetries can be
measured in different regions of the Dalitz plot. In particular these
asymmetries can be measured both along resonances and in the non-resonant
regions. A quantitative estimate  of the variation of these asymmetries obtained
experimentally would provide valuable understanding of $\SU(3)$ breaking
effects. It would also be interesting to find regions of the Dalitz plots where
$\SU(3)$ is a good symmetry. The $\SU(3)$ flavor symmetry subsumes three
important and non-commuting $\SU(2)$ symmetries: isospin (or $T$-spin), $U$-spin
and $V$-spin. All the members of a $\SU(3)$ multiplet are related to one another
by combined operations of the raising and lowering operators of these individual
$\SU(2)$ symmetries. In this paper we restrict ourselves to the breaking of
these $\SU(2)$ symmetries in combinations.

\begin{figure}[t]
\centering \includegraphics[scale=1]{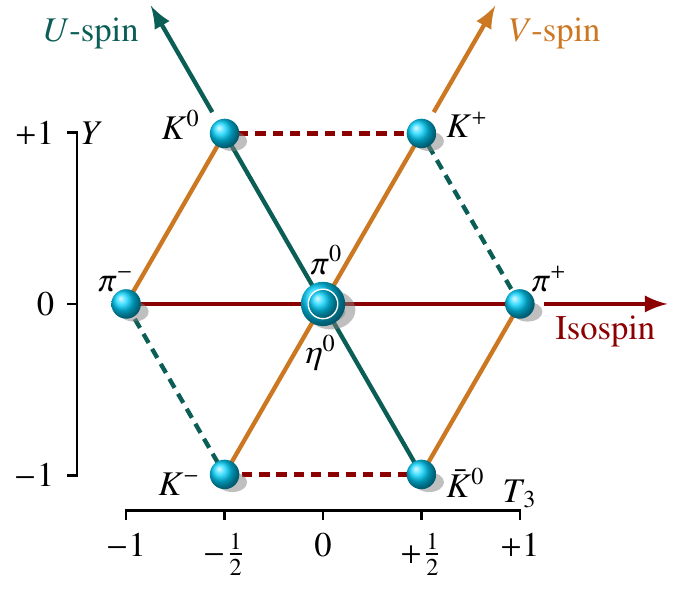} \caption{The
$\SU(3)$ meson octet of light pseudo-scalar mesons. Here the horizontal axis
shows the eigenvalues of isospin ($T_3$) and the vertical axis shows the
eigenvalues of hypercharge ($Y = B+S$, with $B$ being baryon number and $S$
being the strangeness number). The dotted lines parallel to $U$-spin (or
isospin) axis signify that in no two-body decays of $B$ or $D$ meson can the two
connected mesons appear together in the final state as that would violate
conservation of electric charge (or strangeness by two units).}
\label{fig:SU3-meson-octet}
\end{figure}

We shall work with the three-body decays of the type $P \to M_1 M_2 M_3$, where
$P$ can be either a $B$ or a $D$ meson and the final particles $M_1$, $M_2$ and
$M_3$ are distinct members of the lightest pseudo-scalar $\SU(3)$ multiplet (see
Fig.~\ref{fig:SU3-meson-octet}). Our approach towards experimental estimation of
breaking of $\SU(3)$ flavor symmetry, primarily looks for violations of two
constituent $\SU(2)$ symmetries. Therefore, our final state would have a pair of
particles in one $\SU(2)$ multiplet and another pair belonging to a different
$\SU(2)$ multiplet. If the $\SU(2)$ symmetry is assumed to be exact, the pairs
of final state particles that are members of the $\SU(2)$ multiplet are
identical bosons in the symmetry limit and must be totally symmetric under
exchange. This implies that if the wave-function is symmetric under $\SU(2)$
exchange it must be even under space exchange, whereas if it is anti-symmetric
in $\SU(2)$, it must be odd under space exchange too. We shall explicitly
explore this exchange symmetry to deduce some simple relations that predict a
pattern in the distribution of events in the concerned Dalitz plot. Any
deviation from this predicted Dalitz plot distribution would, therefore,
constitute a test of breaking of $\SU(3)$ flavor symmetry. Dalitz plots have
previously been used in Refs.~\cite{Sinha:2011ky, Sahoo:2013mqa, Sahoo:2014nna}
to extract weak decay amplitudes and to study $\CP$, $\CPT$ and Bose 
symmetry violations. Here we use the Dalitz plot to look for breaking of 
$\SU(3)$ flavor symmetry in a single decay mode.

We start Sec.\ref{sec:method} by explaining briefly in
subsection~\ref{subsec:notation} the kind of Dalitz plot we shall use to
elucidate our method and also set up the notation to be followed thenceforth. We
shall then illustrate the method in full detail in subsection~\ref{subsec:IU} by
considering the decay mode $\Bp \to \Kz\piz\pip$ which tests both isospin and
$U$-spin simultaneously. We show in detail how the exchange
$\piz\leftrightarrow\pip$ under isospin and $\Kz\leftrightarrow\piz$ under
$U$-spin results in a characteristic distribution of events in the Dalitz plot
if both isospin and $U$-spin are exact symmetries. The method can equally well
be applied to the decay mode $\Dp_s \to \Kz\piz\pip$. We then show how
$G$-parity generalized to $V$-spin further influences the distribution of events
in the Dalitz plot. The definition of $G$-parity  and its generalization to
$U$-spin and $V$-spin are discussed in the Appendix~\ref{sec:appendix} for ready
reference.  We provide Dalitz plot asymmetries which can then be easily used to
make quantitative estimate of the breaking of $\SU(3)$ flavor symmetry. Then, we
sketch out the necessary steps for handling cases of both isospin and $V$-spin
violation (in subsection~\ref{subsec:IV}) as well as both $U$-spin and $V$-spin
violation (in subsection~\ref{subsec:UV}) by considering the decay modes $\Bz_d
\text{ or } \Bzb_s \to \Kp\piz\pim$   and $\Bp \text{ or } \Dp \to \Kp \piz\Kzb$
respectively. Finally in subsection~\ref{subsec:IUV}, we sketch out as to how
our method can be applied to a decay mode $\Dp \to \pip\piz\Kzb$ where each pair
of particles in the final state can be directly related by one of the three
$\SU(2)$ symmetries, namely isospin, $U$-spin and $V$-spin. We point out how the
Dalitz plot distribution for this mode differs from the ones considered in the
earlier subsections. Finally, we conclude in section~\ref{sec:conclusions}
emphasizing the salient features of our method.

\section{The method} 
\label{sec:method}

\begin{table*}[t]
\centering
\begin{tabular}{|c|c|c|c|c|c|c|}\hline
\multicolumn{3}{|c|}{Final state} & \multicolumn{2}{c|}{Kind of $\SU(2)$
exchange} & \multicolumn{2}{c|}{Expression for the state} \\ \hline %
$M_1$ & $M_2$ & $M_3$ & $M_1 \leftrightarrow M_2$ & $M_2 \leftrightarrow M_3$ &
$\ket{M_1 M_2}$ & $\ket{M_2 M_3}$ \\ \hline \hline %
$\Kz$ & $\piz$ & $\pip$ & $U$-spin & Isospin & $\frac{1}{2\sqrt{2}} \Big(
\ket{2,+1}_U + \ket{1,+1}_U \Big) - \frac{\sqrt{3}}{2} \ket{1',+1}_U$ &
$-\frac{1}{\sqrt{2}} \Big( \ket{2,+1}_I - \ket{1,+1}_I \Big)$ \\ \hline %
$\Kp$ & $\piz$ & $\pim$ & $V$-spin & Isospin & $-\frac{1}{2\sqrt{2}} \Big(
\ket{2,+1}_V + \ket{1,+1}_V \Big) + \frac{\sqrt{3}}{2} \ket{1',+1}_V$ &
$\frac{1}{\sqrt{2}} \Big( \ket{2,-1}_I + \ket{1,-1}_I \Big)$\\ \hline %
$\Kp$ & $\piz$ & $\Kzb$ & $V$-spin & $U$-spin & $-\frac{1}{2\sqrt{2}} \Big(
\ket{2,+1}_V + \ket{1,+1}_V \Big) + \frac{\sqrt{3}}{2} \ket{1',+1}_V$ &
$\frac{1}{2\sqrt{2}} \Big( \ket{2,-1}_U + \ket{1,-1}_U \Big) -
\frac{\sqrt{3}}{2} \ket{1',-1}_U$\\ \hline %
$\pip$ & $\piz$ & $\Kzb$ & Isospin & $U$-spin & $-\frac{1}{\sqrt{2}} \Big(
\ket{2,+1}_I + \ket{1,+1}_I \Big)$ & $-\frac{1}{2\sqrt{2}} \Big( \ket{2,-1}_U +
\ket{1,-1}_U \Big) + \frac{\sqrt{3}}{2} \ket{1',-1}_U$ \\ \hline %
\end{tabular}
\caption{We look at decays with the final states $M_1 M_2 M_3$ given as in the
table here. The particle $M_2$, which is always $\piz$, being at the center of
the pseudoscalar meson octet belongs to all the three $\SU(2)$ symmetries under
consideration. The states are denoted with subscripts for clarity, e.g.\ the 
state $\ket{U=1, U_3 = +1}$ is denoted as
$\ket{1,+1}_U$. Modes with conjugate final states can as well be studied in a
similar manner. The primed states such as $\ket{1',\pm 1}$ arise from the
$\ket{0,0}$ component of $\piz$ under $U$-spin and $V$-spin considerations as
discussed in the text. The last mode in the table with final state
$\pip\piz\Kzb$ has another exchange symmetry, namely exchange of $\pip$ and
$\Kzb$ under $V$-spin. Thus $\ket{\pip\Kzb} = \frac{1}{\sqrt{2}} \Big(
\ket{1,0}_V + \ket{0,0}_V \Big)$ under $V$-spin.} \label{tab:modes}
\end{table*}

\subsection{General considerations}
\label{subsec:notation}

The method described in this paper relies on the simultaneous application of two
of the $\SU(2)$ symmetries subsumed in $\SU(3)$ i.e. isospin (or $T$-spin),
$U$-spin or $V$-spin, to a three body decay $P \to M_1 M_2 M_3$, where $M_1$,
$M_2$ and $M_3$ are chosen such that $M_1$ and $M_2$  belong to the triplet of
one of the $\SU(2)$ subgroups and $M_2$ and $M_3$ belongs to another. To be
definite $M_2$ is always chosen to be the $\piz$ and the modes we consider are
listed in Table~\ref*{tab:modes}.  Under the limit of exact $\SU(2)$ all the
meson belonging to the triplet are identical bosons and must exhibit an overall
Bose symmetry under exchange. This behavior must also be reflected in the Dalitz
plot for the decay. We can construct a Dalitz plot out of the Mandelstam-like
variables $s$, $t$ and $u$. Let us denote the 4-momenta of particles $P$ and
$M_i$ (where $i \in \{1,2,3\}$) by $p$ and $p_i$ and their masses by $m$ and
$m_i$ respectively. The variables $s,t,u$ are defined in terms of the 4-momenta
as follows:
\begin{equation}\label{eq:stu}
\begin{aligned}
s &= \left( p - p_1 \right)^2 = \left( p_2 + p_3 \right)^2,\\
t &= \left( p - p_2 \right)^2 = \left( p_1 + p_3 \right)^2,\\
u &= \left( p - p_3 \right)^2 = \left( p_1 + p_2 \right)^2.
\end{aligned}
\end{equation}
It is easy to observe that $(m_2 + m_3)^2 \leqslant s \leqslant (m - m_1)^2$,
$(m_1 + m_3)^2 \leqslant t \leqslant (m - m_2)^2$, $(m_1 + m_2)^2 \leqslant u
\leqslant (m - m_3)^2$, and $s + t + u = m^2 + m_1^2 + m_2^2 + m_3^2 = M^2 $
(say). 
\begin{figure}[htp]
\centering \includegraphics[width=0.95\linewidth]{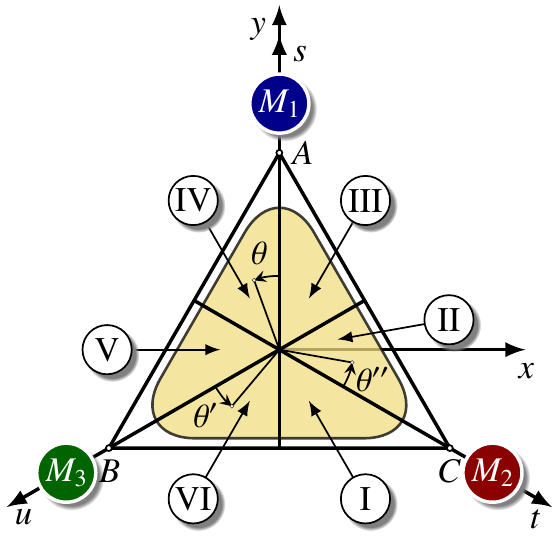} %
\caption{A hypothetical Dalitz plot for the decay $P \to M_1 M_2 M_3$, where the
variables $s,t,u$ are defined in Eq.\eqref{eq:stu}. The three sides of the
equilateral triangle are given by $s=0$, $t=0$ and $u=0$. The three vertices
$A$, $B$, $C$, correspond to $s=M^2$, $u=M^2$ and $t=M^2$ respectively. The
three medians divide the interior of the equilateral triangle into six regions
of equal area. These six sextants are denoted by $I,II,III,IV,V$ and $VI$. The
three vertices $A,B,C$ of the equilateral triangle have rectangular coordinates
$A = (0,2)$, $B=(-\sqrt{3},-1)$ and $C = (\sqrt{3},-1)$. This rectangular
coordinate system has its origin at the center of the equilateral triangle and
the $y$-axis is along the $s$-axis as shown here. The angles $\theta$, $\theta'$
and $\theta''$ are defined in the text.The blobs with $M_1$, $M_2$ and $M_3$
serve as mnemonic to suggest that the exchanges $s \leftrightarrow t
\leftrightarrow u$ are equivalent to the particle exchanges $M_1 \leftrightarrow
M_2 \leftrightarrow M_3$ respectively. The physically allowed region is always
inside the equilateral triangle as shown, schematically, by the yellow colored
region.} %
\label{fig:Dalitz}
\end{figure}
In order to give equal weightage to $s,t$ and $u$ we shall work with a ternary
plot of which $s,t,u$ form the three axes. This leads to an equilateral triangle
as shown in Fig.~\ref{fig:Dalitz}. When the final particles are
ultra-relativistic, the full interior of the equilateral triangle tends to get
occupied. In any case the Dalitz plot under our consideration would always lie
inside the equilateral triangle. The physically allowed region is schematically
shown in Fig.~\ref{fig:Dalitz} by the yellow colored region inside the
equilateral triangle. The boundary of the Dalitz plot for a three-body decay
process under consideration would not look symmetric under the exchanges $s
\leftrightarrow t \leftrightarrow u$ due to the breaking of flavor $\SU(3)$
symmetry on account of masses $m_1$, $m_2$ and $m_3$ being different. Any event 
inside the Dalitz plot, as illustrated in
Fig.~\ref{fig:Dalitz}, can be specified by its radial distance ($r$) from the
center of the equilateral triangle and the angle subtended by its position
vector with any of the three axes $s,t$, or $u$. The angle subtended by the
position vector with $s$-axis is denoted by $\theta$, the one with $u$-axis is
denoted by $\theta'$ and the one with $t$-axis is denoted by $\theta''$. An
event described by some values of $s$, $t$ and $u$ corresponds to some values of
$r$ and $\theta$ as calculable from the relations given below:
\begin{align}
\label{eq:s}
s &= \frac{M^2}{3} \Big( 1 + r \, \cos\theta \Big),\\
\label{eq:t}
t &= \frac{M^2}{3} \left( 1 + r \, \cos\left( \frac{2\pi}{3} + \theta
\right) \right),\\
\label{eq:u}
u &= \frac{M^2}{3} \left( 1 + r \, \cos\left( \frac{2\pi}{3} - \theta
\right) \right).
\end{align}
One can easily change the basis from $(r,\theta)$ to either $(r,\theta')$ or
$(r,\theta'')$ by noting the fact that $\theta = \theta' + \frac{2\pi}{3}$ and
$\theta = \theta'' + \frac{4\pi}{3}$ (see Fig.~\ref{fig:Dalitz}).

Before we analyze the specific decay modes, we would like to point out a few
simple facts about the neutral pion, which plays a pivotal role in all our
decays. The neutral pion is a pure isotriplet state $\ket{1,0}_I \equiv
\frac{1}{\sqrt{2}} \left( d\bar d - u\bar u \right)$:
\begin{equation}
\ket{\piz} = - \ket{1,0}_I.
\end{equation}
But in case of $U$-spin it is a linear combination of the $U$-spin triplet state
$\ket{1,0}_U \equiv \frac{1}{\sqrt{2}} \left( s\bar s - d \bar d \right)$ and
the $U$-spin singlet but $\SU(3)$ octet state $\ket{0,0}_{U,8} \equiv
\frac{1}{\sqrt{6}} \left( d \bar d + s\bar s - 2 u \bar u \right)$:
\begin{equation}
\ket{\piz} = \frac{1}{2} \ket{1,0}_U - \frac{\sqrt{3}}{2} \ket{0,0}_{U,8}.
\end{equation}
Similarly in case of $V$-spin, $\piz$ is given by a linear combination of the
$V$-spin triplet state $\ket{1,0}_V \equiv \frac{1}{\sqrt{2}} \left( s\bar s - u
\bar u \right)$ and the $V$-spin singlet but $\SU(3)$ octet state
$\ket{0,0}_{V,8} \equiv \frac{1}{\sqrt{6}} \left( u \bar u + s\bar s - 2 d \bar
d \right)$:
\begin{equation}
\ket{\piz} = -\frac{1}{2} \ket{1,0}_V + \frac{\sqrt{3}}{2} \ket{0,0}_{V,8}.
\end{equation}
We have put subscripts $I,U,V$ in the states to indicate that they are written
in isospin, $U$-spin and $V$-spin bases respectively.

\subsection{Decay Mode with final state \texorpdfstring{$\Kz\piz\pip$}{K0 pi0
pi+}}\label{subsec:IU}

We begin by considering as an example the decay mode $\Bp \to \Kz\piz\pip$. We
will see that the application of both isospin and $U$-spin results in unique
tests of the validity of both these constituent symmetries of $\SU(3)$.
The $\piz$ and $\pip$ in the final state are identical under isospin and the
final state must be totally symmetric under exchange. Under  $U$-spin (see
Fig.~\ref{fig:SU3-meson-octet}) the $\Kz$ and $\piz$ can be considered as
identical bosons and must similarly be totally symmetric under exchange. This
ensures the following exchanges in the Dalitz plot:
\begin{align*}
& U\text{-spin exchange} \equiv \Kz \leftrightarrow \piz \implies s
\leftrightarrow t,\\
& \text{isospin exchange} \equiv \piz \leftrightarrow \pip \implies t
\leftrightarrow u.
\end{align*}
Under exact $U$-spin and isospin, the final state $\Kz\piz\pip$ has, therefore,
the following two possibilities:
\begin{enumerate}
\item $\Kz\piz$ would exist in either symmetrical or anti-symmetrical state
w.r.t.\ their exchange in space, and %
\item $\piz\pip$ would exist in either symmetrical or anti-symmetrical state
w.r.t.\ their exchange in space.
\end{enumerate}

The amplitude for this decay, would then be described by four independent
functions defined by their symmetry and anti-symmetry properties as enunciated
below:
\begin{enumerate}
\item $\ASS(s,t,u)$ which is symmetric under both $s \leftrightarrow t$ and $t
\leftrightarrow u$, or %
\item $\AAA(s,t,u)$ which is anti-symmetric under both $s \leftrightarrow t$ and
$t \leftrightarrow u$, or %

\item $\ASA(s,t,u)$ which is symmetric under $s \leftrightarrow t$ and
anti-symmetric under $t \leftrightarrow u$, or %
\item $\AAS(s,t,u)$ which is anti-symmetric under $s \leftrightarrow t$ and
symmetric under $t \leftrightarrow u$.
\end{enumerate}

We now analyze each of the possible amplitude functions in the most general
manner. We start by $\ASS(s,t,u)$, which is a function symmetric under both $s
\leftrightarrow t$ and $t \leftrightarrow u$ to show that  $\ASS(s,t,u)$ must
also be symmetric under $s\leftrightarrow u$:
\begin{align*}
\ASS(s,t,u) &\eqst \ASS(t,s,u) \eqtu \ASS(u,s,t) \\
&\eqst \ASS(u,t,s).
\end{align*}
Since, we have shown that $\ASS(s,t,u) = \ASS(u,t,s)$, we have demonstrated that
$\ASS(s,t,u)$  is also symmetric under $s\leftrightarrow u$. Hence, we conclude
that $\ASS(s,t,u)$ is a fully symmetric amplitude function. Let us next consider
$\AAA(s,t,u)$ which is a function anti-symmetric under both $s \leftrightarrow
t$ and $t \leftrightarrow u$ to show that it is also anti-symmetric under
$s\leftrightarrow u$:
\begin{align*} 
\AAA(s,t,u) &\eqst -\AAA(t,s,u) \eqtu +\AAA(u,s,t) \\
& \eqst -\AAA(u,t,s).
\end{align*}
Since, $\AAA(s,t,u) = -\AAA(u,t,s)$ we require that $\AAA(s,t,u)$  must also be
anti-symmetric under $s\leftrightarrow u$. Hence, we conclude that $\AAA(s,t,u)$
is a fully anti-symmetric amplitude function. Following the same arguments as
above it is easy to conclude that both $\ASA(s,t,u)$ and $\AAS(s,t,u)$ must be
identically zero. The details are as follows. The function $\ASA(s,t,u)$ which
is symmetric under $s \leftrightarrow t$ and anti-symmetric under $t
\leftrightarrow u$ must satisfy
\begin{align*}
\ASA(s,t,u) &\eqst \ASA(t,s,u) \eqtu - \ASA(u,s,t) \\
& \eqst -\ASA (u,t,s) \eqtu + \ASA (t,u,s) \\
& \eqst + \ASA (s,u,t) \eqtu - \ASA (s,t,u)=0.
\end{align*}
Similarly, $\AAS(s,t,u)$ being a function anti-symmetric under $s
\leftrightarrow t$ and symmetric under $t \leftrightarrow u$ satisfies
\begin{align*}
\AAS(s,t,u) &\eqst - \AAS(t,s,u) \eqtu - \AAS(u,s,t) \\
& \eqst + \AAS(u,t,s) \eqtu + \AAS (t,u,s) \\
& \eqst - \AAS (s,u,t) \eqtu - \AAS (s,t,u)= 0.
\end{align*}
We have shown that both $\ASA(s,t,u)=0$ and $\ASA(s,t,u)=0$, which implies that
these amplitudes never contribute to the distribution of events on the Dalitz
plot. Since, the function describing the distribution of events in the Dalitz
plot is proportional to the amplitude mod-square, it also has only two parts,
one which is fully symmetric under $s \leftrightarrow t \leftrightarrow u$, and
another which is fully anti-symmetric under $s \leftrightarrow t \leftrightarrow
u$.

We now examine the decay mode $\Bp \to \Kz\piz\pip$ in detail, by writing down
the decay amplitude in terms of isospin and  $U$-spin amplitudes, eventually
obtaining the same conclusion as above about the distribution of events in the
Dalitz plot under consideration. The $\piz\pip$ combination can exist in isospin
states $\ket{2,+1}_I$ and $\ket{1,+1}_I$ (see Table~\ref{tab:modes}). If isospin
were an exact symmetry, the state $\ket{\piz\pip}$ would remain unchanged under
$\piz \leftrightarrow \pip$ exchange. This puts the $\ket{2,+1}_I$ state in a
space symmetric (even partial wave) state, and the $\ket{1,+1}_I$ state in a
space anti-symmetric (odd partial wave) state. The isospin decomposition of the
final state $\ket{\Kz\piz\pip}$ is given by
\begin{align}
\ket{\Kz\piz\pip} &= -\frac{1}{\sqrt{5}} \ket{\frac{5}{2},+\frac{1}{2}}_I^e +
\frac{\sqrt{3}}{\sqrt{10}} \ket{\frac{3}{2}, + \frac{1}{2}}_I^e \nn\\%
& \quad + \frac{1}{\sqrt{6}} \ket{\frac{3}{2}, + \frac{1}{2}}_I^o - \frac{1}{\sqrt{3}} \ket{\frac{1}{2}, +\frac{1}{2}}_I^o,
\end{align}
where the superscripts $e,o$ denote the even, odd nature of the state under the
exchange $\piz \leftrightarrow \pip$. The sign change in the odd states above is
due to the odd $\ket{1,+1}_I$ isospin component of the $\ket{\piz\pip}$ state
switching sign under $\piz \leftrightarrow \pip$ exchange, whereas the
$\ket{2,+1}_I$ is even under the same exchange. Since $\Bp$ has isospin state
$\ket{\frac{1}{2},+\frac{1}{2}}_I$, and only $\Delta I = 0,1$ currents are
allowed by the Hamiltonian in standard model, we would have no contributions
from $\ket{\frac{5}{2}, + \frac{1}{2}}_I$ state. The $\ket{\frac{3}{2}, +
\frac{1}{2}}_I$ state can arise from both $\ket{\frac{1}{2}, - \frac{1}{2}}_I
\otimes \ket{2,+1}_I$ and $\ket{\frac{1}{2}, - \frac{1}{2}}_I \otimes
\ket{1,+1}_I$, with the first contribution being symmetric and the later being
anti-symmetric. The state $\ket{\frac{1}{2}, + \frac{1}{2}}_I$ on the other-hand
is purely anti-symmetric. Even though we shall work with the standard model
Hamiltonian, our conclusions are general and are valid even when more general
Hamiltonians exist.

The isospin $I=\frac{1}{2}$ initial state decays to a final state that can be
decomposed into either $I=\frac{1}{2}$ or $I=\frac{3}{2}$ states via a
Hamiltonian that allows $\Delta I=0$ or $\Delta I=1$ transitions. The transition
with $\Delta I=1$ results in two amplitudes with $I=\frac{1}{2}$ or
$I=\frac{3}{2}$ represented as $\T{1}$ and $\T{3}$ respectively, whereas $\Delta
I=0$ transition results only in a single amplitude with final state
$I=\frac{1}{2}$ labeled as $\t$. The isospin amplitudes $\T{1}$, $\T{3}$ and
$\t$ are themselves defined~\cite{Lipkin:1991st} in terms of the Hamiltonian to
be:
\begin{eqnarray}
  \label{eq:T} %
  \T{3}&=&\sqrt{\frac{1}{3}} \bracket{\frac{3}{2},\pm\frac{1}{2}}{\mathscr{
  H}_{\Delta I=1}}{\frac{1}{2},\pm\frac{1}{2}},\nn \\ %
  \T{1}&=&\pm\sqrt{\frac{2}{3}}\bracket{\frac{1}{2},\pm\frac{1}{2}}{\mathscr{
  H}_{\Delta I=1}}{\frac{1}{2},\pm\frac{1}{2}},\nn \\ %
  \t&=&\sqrt{\frac{2}{3}}\bracket{\frac{1}{2},\pm\frac{1}{2}}{\mathscr{
  H}_{\Delta I=0}}{\frac{1}{2},\pm\frac{1}{2}}.
\end{eqnarray}
The amplitude for the decay $\Bp \to \Kz\piz\pip$  can then be written in terms 
of the isospin amplitudes as 
\begin{align}
\label{eq:Tamp} A(B^+\to K^0\pi^0 \pi^+)&=\dsp\frac{3}{\sqrt{10}} \T{3}^{\e}\,X
\nn\\ %
& \quad - \frac{1}{\sqrt{2}}  \Big( \T{3}^{\o} + \T{1}^{\o}+ \t^{\o}
\Big)\,Y\sin\theta,
\end{align}
where $X$ and $Y\sin\theta$ are introduced to take care of the spatial and
kinematic contributions as is seen from the discussion above (see
Eqns.~\eqref{eq:t} and \eqref{eq:u}). In general, $X$ and $Y$ can be arbitrary
functions of $r$ and $\cos\theta$.  The functions $X$ and $Y$ are in general
mode dependent, however, they are same for modes related by isospin symmetry. We
retain the subscripts `$\e$' and `$\o$' merely to track the even or odd isospin
state of the two pion in the three-body final state.

On the other hand, if $U$-spin were an exact symmetry the state $\Kz\piz$ must
remain unchanged under $\Kz \leftrightarrow \piz$ exchange. Under $U$-spin the
$\Kz\piz$ state can exist in $\ket{2,+1}_U$ and $\ket{1,+1}_U$ (see
Table~\ref{tab:modes}), out of which $\ket{1,+1}_U$ has a contribution from the
$\ket{0,0}_{U,8}$ admixture in $\piz$ which is denoted by $\ket{1',+1}_U$. Both
$\ket{2,+1}_U$ and the $\ket{1,+1}_U$ coming from the $\ket{0,0}_{U,8}$
contribution of $\piz$ exist in space symmetric (even partial wave) states, and
that part of $\ket{1,+1}_U$ arising out of $\ket{1,0}_U$ part of $\piz$ exists
in space anti-symmetric (odd partial wave) state. The $U$-spin decomposition of
the final state $\ket{\Kz\piz\pip}$ is given by
\begin{align}
\ket{\Kz\piz\pip} &= -\frac{1}{2\sqrt{5}} \ket{\frac{5}{2},+\half}_U^e -
\frac{\sqrt{3}}{2\sqrt{10}}\ket{\frac{3}{2},+\half}_U^e \nn\\ %
& \quad - \frac{1}{2\sqrt{6}} \ket{\frac{3}{2},+\half}_U^o - \frac{1}{2\sqrt{3}}
\ket{\half, +\half}_U^o \nn\\ %
& \quad + \frac{1}{2} \ket{\frac{3'}{2},+\half}_U^e + \frac{1}{\sqrt{2}}
\ket{\frac{1'}{2}, +\half}_U^e,
\end{align}
where the superscripts $e,o$ denote that the state is even, odd under the
exchange $\Kz \leftrightarrow \piz$. The origin of sign change in the odd terms
above is easy to understand from the $U$-spin decomposition of the
$\ket{\Kz\piz}$ state:
\begin{equation*}
\ket{\Kz\piz} = \frac{1}{2\sqrt{2}} \Big( \ket{2,+1}_U + \ket{1,+1}_U \Big) - 
\frac{\sqrt{3}}{2} \ket{1',+1}_U ,
\end{equation*}
which transforms as follows under the $\Kz \leftrightarrow \piz$ exchange
\begin{equation*}
\ket{\piz\Kz}=\frac{1}{2\sqrt{2}} \Big( \ket{2,+1}_U - \ket{1,+1}_U \Big) - 
\frac{\sqrt{3}}{2} \ket{1',+1}_U .
\end{equation*}
We recollect that $\ket{1,+1}_U$ is an odd state under $\Kz \leftrightarrow
\piz$ exchange, whereas $\ket{2,+1}_U$ and $\ket{1',+1}_U$ are even states under
the same exchange. It is easy to see that $\ket{\frac{5}{2}, + \frac{1}{2}}_U$
and $\ket{\frac{3}{2}, + \frac{1}{2}}_U$ states do not contribute since the
parent particle $\Bp$ is a $U$-spin singlet, and only the  $\Delta U =
\tfrac{1}{2}$ current contributes to the decay. This unique feature follows from
the fact that the electroweak penguin does not violate $U$-spin as $d$ and $s$
quarks carry the same electric charge (see \cite{Soni:2006vi}). Hence, only the
$\ket{\frac{1}{2}, \frac{1}{2}}_U$ and $\ket{\frac{1'}{2}, \frac{1}{2}}_U$ can
contribute to the decay amplitude and they correspond  to anti-symmetric and
symmetric contributions under $\Kz\leftrightarrow\piz$ respectively. The
$U$-spin amplitudes
\begin{eqnarray}
  \label{eq:U}
  \u&=&\pm\sqrt{\frac{2}{3}}\biggbracket{\frac{1}{2},\pm\frac{1}{2}}{\mathscr{
  H}_{\Delta U=\tfrac{1}{2}}}{0,0}~,\nn \\ %
  \up{} &=&\sqrt{\frac{1}{3}}\biggbracket{\frac{1'}{2},\pm\frac{1}{2}}{\mathscr{
  H}_{\Delta U=\tfrac{1}{2}}}{0,0}.
\end{eqnarray}
Hence, the amplitude for the decay $\Bp \to \Kz\piz\pip$  can then be written in
terms of the $U$-spin amplitudes as
\begin{equation}
\label{eq:Uamp}
A(B^+\to K^0\piz \pip)=\dsp\frac{3}{\sqrt{10}} \up{\e}  \,X^\prime +\u^{\o} 
\,Y^\prime\sin\theta^\prime, 
\end{equation}
where $X^\prime$ and $Y^\prime$ are functions that are, in general, arbitrary
functions of $r$ and $\cos\theta'$, and are introduced to take care of spatial
and kinematic contributions to the decay amplitude. The subscripts `$\e$' and
`$\o$' are again retained to merely track the even or odd $U$-spin state of
$\Kz$ and $\piz$ in the three-body final state. As argued earlier the amplitude
for the decay has two parts, one fully symmetric under the exchanges $s
\leftrightarrow t \leftrightarrow u$ (i.e.\ $\ASS(s,t,u)$) and another fully
anti-symmetric under the same exchanges (i.e.\ $\AAA(s,t,u)$). Comparing
Eqs.~\eqref{eq:Tamp} and \eqref{eq:Uamp} we obtain:
\begin{eqnarray}
\label{eq:amp-even} \ASS &=& \frac{3}{\sqrt{10}} \T{3}^{\e}\,X
=\frac{3}{\sqrt{10}} \up{\e} \,X^\prime\\ %
\label{eq:amp-odd} \AAA &=& - \frac{1}{\sqrt{2}} \Big( \T{3}^{\o} + \T{1}^{\o} +
\t^{\o} \Big) \,Y\sin\theta \nn \\&=&\u^{\o} \,Y^\prime\sin\theta^\prime.
\end{eqnarray}
The exchange $s \leftrightarrow t \leftrightarrow u$ being equivalent to $\theta
\leftrightarrow \theta' \leftrightarrow \theta''$, implies that the fully
anti-symmetric amplitude $\AAA(s,t,u)$ must be proportional to $\sin3\theta$
because $\sin3\theta = \sin3\theta'=\sin3\theta''$ as $\theta = \theta' +
\frac{2\pi}{3} = \theta'' + \frac{4\pi}{3}$. From elementary trigonometry we
know that $\sin3\theta = \sin\theta \, \left( 4 \cos^2\theta -1 \right)$. This
implies that the factor $\left(4 \cos^2\theta -1\right)$ is an even function of
$\cos\theta$ and is a part of both $Y$ and $Y'$ in Eq.~\eqref{eq:amp-odd}, i.e.\
$ Y = y \left( 4 \cos^2\theta -1 \right)$ and $Y' = y' \left( 4 \cos^2\theta' -1
\right)$ for some $y$ and $y'$ such that
\begin{align}
\label{eq:3theta}
\AAA &= - \frac{1}{\sqrt{2}} \Big( \T{3}^{\o} + \T{1}^{\o}+
\t^{\o} \Big) \,y\sin3\theta \nn \\
&=\u^{\o} \,y'\sin3\theta^\prime.
\end{align}

\begin{figure}[hbtp]
\centering \includegraphics[width=0.95\linewidth]{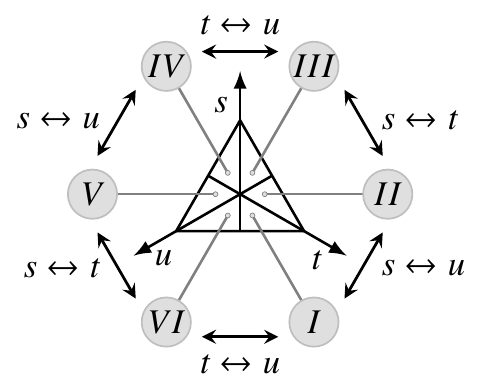}
\caption{Exchanges that take us from one sextant to another in the Dalitz plot.
It must be noted that the following exchanges are also equivalent: $s
\leftrightarrow t \leftrightarrow u \equiv \theta \leftrightarrow \theta'
\leftrightarrow \theta''$ as well as $t \leftrightarrow u \equiv \theta
\leftrightarrow -\theta$, $s \leftrightarrow t \equiv \theta' \leftrightarrow
-\theta'$ and $u \leftrightarrow s \equiv \theta'' \leftrightarrow -\theta''$.}
\label{fig:Dalitz_plot_exhange}
\end{figure}

The Dalitz plot can be divided into six sextants by means of the $s$, $t$ and
$u$ axes which go along the medians of an equilateral triangle as shown in
Figs.~\ref{fig:Dalitz} and \ref{fig:Dalitz_plot_exhange}. Since the Dalitz plot
distribution function is proportional to the amplitude mod-square, it would also
have a part which is fully symmetric under $s \leftrightarrow t \leftrightarrow
u$ (denoted by $f_{SS} (s,t,u)$) and another part which is fully anti-symmetric
under the same exchanges (denoted by $f_{AA}(s,t,u)$):
\begin{align}
\label{eq:fss}
f_{SS}(s,t,u) &\propto \modulus{\ASS(s,t,u)}^2 + \modulus{\AAA(s,t,u)}^2,\\
\label{eq:faa}
f_{AA}(s,t,u) &\propto 2 \, \Re\Big(\ASS(s,t,u) \cdot \AAA^*(s,t,u) \Big).
\end{align}
Let us denote the function describing distribution of events in any sextant, say
the $i$th one, of the Dalitz plot by $f_i (r,\theta)$, where the coordinates
$(r,\theta)$ lie in the sextant $i$ and we could have as well used the other
equivalent choices $\theta'$ or $\theta''$ instead of $\theta$, the choice of
which is subject to the underlying symmetry being considered (see
Fig.~\ref{fig:Dalitz_plot_exhange}). Henceforth we shall drop $(r,\theta)$ from
the distribution functions, except when necessary, as we implicitly assume the
$r$ and $\theta$ dependence in them. The distribution function must have only
two parts as said above, the fully symmetric and the fully anti-symmetric parts.
Let us assume that in sextant $I$ the Dalitz plot distribution is given by the
function
\begin{equation}
f_{I} = f_{SS} (s,t,u) + f_{AA} (s,t,u).
\end{equation}
It is then trivial to see that the Dalitz plot distributions in the even
numbered sextants should be identical to one another, and the odd numbered
sextants would also be identically populated, because
\begin{align}
\label{eq:fodd}
f_{I} = f_{III} = f_{V} &= f_{SS} (s,t,u) + f_{AA} (s,t,u),\\
\label{eq:feven}
f_{II} = f_{IV} = f_{VI} &= f_{SS} (s,t,u) - f_{AA} (s,t,u).
\end{align}
This is the signature of exact $\SU(3)$ flavor symmetry in the Dalitz plots
under our consideration. Any deviation from this conclusion would constitute an
observable evidence for violation of the $\SU(3)$ flavor symmetry.

Until now the exchange properties of $\Kz\leftrightarrow \piz$ under $U$-spin
and $\piz\leftrightarrow \pip$ under isospin have been used to obtain the even
and odd amplitudes contributing to $\Bp\to \Kz\piz\pip$. Since $\Kz$ and $\pip$
belong to different multiplets of $V$-spin, in order to consider the symmetry
properties under $\Kz\leftrightarrow \pip$ one needs to define the $G$-parity
analogue of $V$-spin, denoted by $\GV$ and defined in the 
Appendix~\ref{sec:appendix}. Since charge
conjugation is a good symmetry in strong interaction, $\GV$ is as good as
$V$-spin itself. The state $\ket{\Kz\pip}$ is composed of states which are even
and odd under $\GV$-parity:
\begin{equation*}
\ket{\Kz\pip} = \frac{1}{2} \left( \ket{\Kz\pip}_\e + \ket{\Kz\pip}_\o
\right),
\end{equation*}
where 
\begin{align*}
\ket{\Kz\pip}_\e &= \ket{\Kz\pip} - \ket{\pip\Kz},\\
\ket{\Kz\pip}_\o &= \ket{\Kz\pip} + \ket{\pip\Kz},
\end{align*}
and
\begin{align*}
\GV \ket{\Kz\pip}_\e &= + \ket{\Kz\pip}_\e,\\
\GV \ket{\Kz\pip}_\o &= - \ket{\Kz\pip}_\o.
\end{align*}
We have already proven that the amplitudes for the decay $\Bp\to\Kz\piz\pip$ has
two parts one even and the other odd under the exchange of any two particles in
the final state. Hence, $\ASS$ is odd under $\GV$ and $\AAA$ is even under
$\GV$. Since the two $\GV$-parity amplitudes do not interfere the two amplitudes
$\ASS$ and $\AAA$ do not interfere in the Dalitz plot distribution resulting in
$f_{AA}$ being zero (Eq.~\eqref{eq:faa}). Therefore if $\GV$ is a good symmetry 
of nature  it is interesting to conclude that the Dalitz plot is completely
symmetric under $s\leftrightarrow t \leftrightarrow u$. This implies that
\begin{align}
f_{I} =f_{II} = f_{III} = f_{IV} = f_{V}= f_{VI} \equiv f_{SS} (s,t,u) .
\end{align}
This expression holds only if isospin, $U$-spin and $V$-spin are all exact
symmetries.  However, if $\GV$ is broken the Dalitz plot distribution will still
follow Eqs.~\eqref{eq:fodd} and \eqref{eq:feven} when isospin and $U$-spin are
exact symmetries. In the case when $\GV$ is exact, the exchange properties of
the distribution functions $f_{I}$ to $f_{VI}$ imply that if, (a) $U$-spin is an
exact symmetry, then $f_{II}=f_{III}$, $f_{I}=f_{IV}$ and $f_{V}=f_{VI}$
irrespective of the validity of isospin symmetry, (b) isospin is an exact
symmetry, then $f_{II}=f_{V}$, $f_{I}=f_{VI}$ and $f_{III}=f_{IV}$ irrespective
of the validity of $U$-spin symmetry.
However, when both $\GV$ and either isospin or $U$-spin is broken, then the 
Eqs.~\eqref{eq:fodd} and \eqref{eq:feven} are no longer valid. In such a case, 
we have the following possibilities:
\begin{itemize}
\item \underline{Test for isospin symmetry:} By isospin symmetry, the sextants
$I,II,III$ get mapped to the sextants $VI,V,IV$ respectively. We note that when
isospin is \textit{not} broken, then
\begin{align}
f_{I} + f_{VI} &= f_{III} + f_{IV} = f_{V} + f_{II} = 2 f_{SS} (s,t,u),\\
f_{I} - f_{VI} &= f_{III} - f_{IV} = f_{V} - f_{II} = 2 f_{AA} (s,t,u).
\end{align}
However, when isospin is broken, the values of $f_{SS}$ and $f_{AA}$ extracted
from sextants $I$ and $VI$ need not be same with those extracted from either
$II$ and $V$ or $III$ and $IV$. For further clarification of this statement, we
introduce two quantities $\Sigma^i_j$ and $\Delta^i_j$ defined as
\begin{align}
\label{eq:Sigma}
\Sigma^i_j (r,\theta) &= f_i + f_j,\\
\label{eq:Delta}
\Delta^i_j (r,\theta) &= f_i - f_j,
\end{align}
where $i$ and $j$ are two sextants and $i\neq j$. For conciseness of
expressions, we shall also drop the explicit $(r,\theta)$ dependence of
$\Sigma^i_j$ and $\Delta^i_j$. In terms of these quantities, the signature of
isospin breaking can be succinctly summarized by the inequalities
\begin{align}
\Sigma^{I}_{VI} \neq \Sigma^{III}_{IV} \neq \Sigma^{V}_{II},\\
\Delta^{I}_{VI} \neq \Delta^{III}_{IV} \neq \Delta^{V}_{II}.
\end{align}
An asymmetry can now the constructed to measure the isospin breaking as follows:
\begin{align}
\label{eq:asyiso}
\mathbb{A}_{\text{Isospin}} &= \modulus{\frac{\Sigma^{I}_{VI} -
\Sigma^{III}_{IV}}{\Sigma^{I}_{VI} + \Sigma^{III}_{IV}}} +
\modulus{\frac{\Sigma^{III}_{IV} - \Sigma^{V}_{II}}{\Sigma^{III}_{IV} +
\Sigma^{V}_{II}}} + \modulus{\frac{\Sigma^{V}_{II} -
\Sigma^{I}_{VI}}{\Sigma^{V}_{II} + \Sigma^{I}_{VI}}} \nn\\ %
& \quad + \modulus{\frac{\Delta^{I}_{VI} - \Delta^{III}_{IV}}{\Delta^{I}_{VI} +
\Delta^{III}_{IV}}} + \modulus{\frac{\Delta^{III}_{IV} -
\Delta^{V}_{II}}{\Delta^{III}_{IV} + \Delta^{V}_{II}}} +
\modulus{\frac{\Delta^{V}_{II} - \Delta^{I}_{VI}}{\Delta^{V}_{II} +
\Delta^{I}_{VI}}}.
\end{align}

\item \underline{Test for $U$-spin symmetry:} By $U$spin symmetry, the sextants
$VI,I,II$ get mapped to the sextants $V,IV,III$ respectively. We note that when
$U$-spin is \textit{not} broken, then
\begin{align}
\Sigma^{I}_{IV} = \Sigma^{III}_{II} = \Sigma^{V}_{VI} = 2 f_{SS}(s,t,u),\\
\Delta^{I}_{IV} = \Delta^{III}_{II} = \Delta^{V}_{VI} = 2 f_{AA}(s,t,u).
\end{align}
Here it is profitable to consider the $\Sigma$'s and $\Delta$'s being functions
of $(r,\theta')$ as we are considering $s \leftrightarrow t $ exchange which is
equivalent to $\theta' \leftrightarrow -\theta'$. When $U$-spin is broken
\begin{align}
\Sigma^{I}_{IV} \neq \Sigma^{III}_{II} \neq \Sigma^{V}_{VI},\\
\Delta^{I}_{IV} \neq \Delta^{III}_{II} \neq \Delta^{V}_{VI}.
\end{align}
The asymmetry for $U$-spin breaking is, therefore, given by
\begin{align}
\label{eq:asyU}
\mathbb{A}_{U\text{-spin}} &= \modulus{\frac{\Sigma^{I}_{IV} -
\Sigma^{III}_{II}}{\Sigma^{I}_{IV} + \Sigma^{III}_{II}}} +
\modulus{\frac{\Sigma^{III}_{II} - \Sigma^{V}_{VI}}{\Sigma^{III}_{II} +
\Sigma^{V}_{VI}}} + \modulus{\frac{\Sigma^{V}_{VI} -
\Sigma^{I}_{IV}}{\Sigma^{V}_{VI} + \Sigma^{I}_{IV}}} \nn\\ %
& \quad + \modulus{\frac{\Delta^{I}_{IV} - \Delta^{III}_{II}}{\Delta^{I}_{IV} +
\Delta^{III}_{II}}} + \modulus{\frac{\Delta^{III}_{II} -
\Delta^{V}_{VI}}{\Delta^{III}_{II} + \Delta^{V}_{VI}}} +
\modulus{\frac{\Delta^{V}_{VI} - \Delta^{I}_{IV}}{\Delta^{V}_{VI} +
\Delta^{I}_{IV}}}.
\end{align}

\item \underline{Test for $V$-spin symmetry:} As said before, $\GV$-parity is as badly broken as the $V$-spin because charge conjugation is a good symmetry. When $V$-spin symmetry is broken, then $\GV$ is also broken, and the distribution of events in the Dalitz plot sextants would follow Eqs.~\eqref{eq:fodd} and \eqref{eq:feven}. In addition to that, when $V$-spin is broken, $\Kz$ and $\pip$ need not be even under exchange. This leads to 
\begin{align}
\Sigma_{IV}^{V} \neq \Sigma_{VI}^{III} \neq \Sigma_{II}^{I},\\
\Delta_{IV}^{V} \neq \Delta_{VI}^{III} \neq \Delta_{II}^{I}.
\end{align}
The asymmetry for $V$-spin breaking is, therefore, given by
\begin{align}
\label{eq:asyV}
\mathbb{A}_{V\text{-spin}} &= \modulus{\frac{\Sigma^{V}_{IV} -
\Sigma^{I}_{II}}{\Sigma^{V}_{IV} + \Sigma^{I}_{II}}} +
\modulus{\frac{\Sigma^{I}_{II} - \Sigma^{III}_{VI}}{\Sigma^{I}_{II} +
\Sigma^{III}_{VI}}} + \modulus{\frac{\Sigma^{III}_{VI} -
\Sigma^{V}_{IV}}{\Sigma^{III}_{VI} + \Sigma^{V}_{IV}}} \nn\\ %
& \quad + \modulus{\frac{\Delta^{V}_{IV} - \Delta^{I}_{II}}{\Delta^{V}_{IV} +
\Delta^{I}_{II}}} + \modulus{\frac{\Delta^{I}_{II} -
\Delta^{III}_{VI}}{\Delta^{I}_{II} + \Delta^{III}_{VI}}} +
\modulus{\frac{\Delta^{III}_{VI} - \Delta^{V}_{IV}}{\Delta^{III}_{VI} +
\Delta^{V}_{IV}}}.
\end{align}
\end{itemize}

Hence, the extent of the breaking of isospin, $U$-spin and $V$-spin can easily
be measured from the Dalitz plot distribution. The asymmetries measuring
isospin, $U$-spin and $V$-spin are functions of $r$ and $3\theta\equiv
3\theta'\equiv 3\theta''$ (see the discussion leading to Eq.~\eqref{eq:3theta}).
These asymmetries are, thus, valid in the full Dalitz plot, including the
resonant contributions and can be measured in different regions of the Dalitz
plot. In particular these asymmetries can be measured both along resonances and
in the non-resonant regions.  A quantitative estimate  of the variation of these
asymmetries obtained experimentally would provide valuable understanding of
$\SU(3)$ breaking effects.  It would also be interesting to find regions of the
Dalitz plots where $\SU(3)$ is a good symmetry. The procedure discussed above
can also be applied to other decay modes with the same final state. In
particular one can study the Dalitz plot distribution for the decay $\Dp_s \to
\Kz\piz\pip$ in a similar manner. The amplitudes for this mode are tabulated in
Table.~\ref{tab:KzPizPip}.

\begin{table*}
\centering
\begin{tabular}{|c|c|c|l||c|c|c|l|}
\hline 
\multicolumn{8}{|c|}{$\Bp \to \Kz\piz\pip$} \\ \hline
\multicolumn{4}{|c||}{Isospin \Big(initial state $\ket{\half,+\half}$\Big)} & 
\multicolumn{4}{c|}{$U$-spin \Big(initial state $\ket{0,0}$\Big)}  \\ 
\hline 
 transition & final state & symmetry & Amplitude &  transition & final state & 
 symmetry & Amplitude \\ 
\hline 
 $\Delta I =1$ & $\ket{\thalf,+\half}$ & mixed & 
 $\frac{3}{\sqrt{10}}T_{\!1,\thalf}^e X + \frac{1}{\sqrt{2}} T_{\!1,\thalf}^o Y 
 \sin\theta$ &  $\Delta U=\half$ & $\ket{\half,+\half}$ & odd & 
 $-\frac{1}{2\sqrt{2}}U_{\!\half,\thalf}^o Y' \sin\theta'$ \\ 
\hline
 $\Delta I =1$ & $\ket{\half, +\half}$ & odd & 
 $-\frac{1}{\sqrt{2}}T_{\!1,\half}^o Y \sin\theta$ &   $\Delta U=\half'$ & 
 $\ket{\half,+\half'}$ & even & 
 $\frac{\sqrt{3}}{\sqrt{2}}U_{\!\half,\half'}^{\prime e} X'$ \\ 
\hline
 $\Delta I =0$ & $\ket{\half,+\half}$ & odd & $-\frac{1}{\sqrt{2}} 
 T_{\!0,\half}^o Y \sin\theta$ &  \multicolumn{4}{c|}{} \\ 
\hline \hline
\multicolumn{8}{|c|}{$\Dp_s \to \Kz\piz\pip$} \\ \hline
\multicolumn{4}{|c||}{Isospin \Big(initial state $\ket{0,0}$\Big)} & 
\multicolumn{4}{c|}{$U$-spin \Big(initial state $\ket{\half,+\half}$\Big)} \\ 
\hline 
 transition & final state & symmetry & Amplitude &  transition & final state & 
 symmetry & Amplitude \\ 
\hline 
 $\Delta I =\thalf$ & $\ket{\thalf,+\half}$ & mixed & 
 $\frac{\sqrt{3}}{\sqrt{10}} T_{\!\thalf,\thalf}^e X + \frac{1}{\sqrt{6}} 
 T_{\!\thalf,\thalf}^o Y \sin\theta$ & 
 $\Delta U=1$ & $\ket{\thalf,+\half}$ & mixed & $-\frac{3}{2\sqrt{10}} 
 U_{1,\thalf}^e X' - \frac{1}{2\sqrt{2}} U_{\!1,\thalf}^o Y' \sin\theta'$ \\ 
\hline
 $\Delta I =\half$ & $\ket{\half, +\half}$ & odd & 
 $-\frac{1}{\sqrt{2}}T_{\!\half,\half}^o Y \sin\theta$ &  $\Delta U=1$ & 
 $\ket{\thalf',+\half}$ & even & $\frac{\sqrt{3}}{2} U_{\!1,\thalf'}^{\prime e} 
 X'$ \\ 
\hline
\multicolumn{4}{|c||}{}   & $\Delta U = 1$ & $\ket{\half,+\half}$ & odd & 
$-\frac{1}{2\sqrt{2}} U_{1,\half}^o Y' \sin\theta'$ \\
\cline{5-8} 
\multicolumn{4}{|c||}{}   & $\Delta U = 1$ & $\ket{\half',+\half}$ & even & 
$\frac{\sqrt{3}}{2} U_{1,\half'}^{\prime e} X'$ \\
\cline{5-8} 
\multicolumn{4}{|c||}{}  & $\Delta U = 0$ & $\ket{\half,+\half}$ & odd & 
$-\frac{1}{2\sqrt{2}} U_{0,\half}^o Y' \sin\theta'$ \\
\cline{5-8} 
\multicolumn{4}{|c||}{} & $\Delta U = 0$ & $\ket{\half',+\half}$ & even & 
$\frac{\sqrt{3}}{2} U_{0,\half'}^{\prime e} X'$ \\
\hline
\end{tabular}
\caption{Comparison of decays of $\Bp$ and $\Dp_s$ to the final state $\Kz\piz\pip$.}
\label{tab:KzPizPip}
\end{table*}

\subsection{Decay Mode with final state \texorpdfstring{$\Kp\piz\pim$}{Kp pi0
pi-}}\label{subsec:IV}

\begin{table*}
\centering
\begin{tabular}{|c|c|c|l||c|c|c|l|}
\hline 
\multicolumn{8}{|c|}{$\Bz_d \to \Kp\piz\pim$} \\ \hline
\multicolumn{4}{|c||}{Isospin \Big(initial state $\ket{\half,-\half}$\Big)} & 
\multicolumn{4}{c|}{$V$-spin \Big(initial state $\ket{0,0}$\Big)} \\ 
\hline 
 transition & final state & symmetry & Amplitude & transition & final state & 
 symmetry & Amplitude \\ 
\hline 
 $\Delta I =1$ & $\ket{\frac{3}{2},-\half}$ & mixed & $\frac{3}{\sqrt{10}} 
 T_{\!1,\frac{3}{2}}^e X + \frac{1}{\sqrt{2}} T_{\!1,\frac{3}{2}}^o Y 
 \sin\theta$ &  $\Delta V=\frac{3}{2}$ & $\ket{\frac{3}{2},+\half}$ & mixed & 
 $\frac{\sqrt{3}}{2\sqrt{10}} V_{\!\thalf,\thalf}^e X'' + \frac{1}{2\sqrt{6}} 
 V_{\!\thalf,\thalf}^o Y'' \sin\theta''$ \\ 
\hline
 $\Delta I =1$ & $\ket{\half, -\half}$ & odd & $-\frac{1}{\sqrt{2}} 
 T_{\!1,\half}^o Y \sin\theta$ &  $\Delta V=\frac{3}{2}$ & 
 $\ket{\frac{3}{2}',+\half}$ & even & $-\half 
 V_{\!\frac{3}{2},\frac{3}{2}'}^{\prime e} X''$ \\ 
\hline
 $\Delta I =0$ & $\ket{\half,-\half}$ & odd & $\frac{1}{\sqrt{2}} 
 T_{\!0,\half}^o Y \sin\theta$ &  $\Delta V= \half$ & $\ket{\half,+\half}$ & 
 odd & $\frac{1}{2\sqrt{2}} V_{\!\half,\half}^o Y'' \sin\theta''$ \\ 
\hline
\multicolumn{4}{|c||}{}  & $\Delta V= \half$ & $\ket{\half',+\half}$ & even & 
$-\frac{\sqrt{3}}{\sqrt{2}} V_{\!\half,\half'}^{\prime e} X''$ \\ 
\hline \hline
\multicolumn{8}{|c|}{$\Bzb_s \to \Kp\piz\pim$} \\ \hline
\multicolumn{4}{|c||}{Isospin \Big(initial state $\ket{0,0}$\Big)} & 
\multicolumn{4}{c|}{$V$-spin \Big(initial state $\ket{\half,+\half}$\Big)} \\ 
\hline 
 transition & final state & symmetry & Amplitude &  transition & final state & 
 symmetry & Amplitude \\ 
\hline 
 $\Delta I =\thalf$ & $\ket{\thalf,-\half}$ & mixed & 
 $\frac{\sqrt{3}}{\sqrt{10}} T_{\!\thalf,\thalf}^e X + \frac{1}{\sqrt{6}} 
 T_{\!\thalf,\thalf}^o Y \sin\theta$ & $\Delta V=1$ & $\ket{\thalf,+\half}$ & 
 mixed & $\frac{3}{2\sqrt{10}} V_{\!1,\thalf}^e X'' + \frac{1}{2\sqrt{2}} 
 V_{\!1,\thalf}^o Y'' \sin\theta''$ \\ 
\hline
  $\Delta I =\half$ & $\ket{\half, -\half}$ & odd & $-\frac{1}{\sqrt{2}} 
  T_{\!\half,\half}^o Y \sin\theta$ & $\Delta V=1$ & 
  $\ket{\frac{3}{2}',+\half}$ & even & $-\frac{\sqrt{3}}{2} 
  V_{\!1,\thalf'}^{\prime e} X''$ \\ 
\hline
\multicolumn{4}{|c||}{}  & $\Delta V=1$ & $\ket{\half,+\half}$ & odd & 
$\frac{1}{2\sqrt{2}} V_{\!1,\half}^o Y'' \sin\theta''$ \\ 
\cline{5-8} 
\multicolumn{4}{|c||}{}  & $\Delta V=1$ & $\ket{\half',+\half}$ & even & 
$-\frac{\sqrt{3}}{2} V_{\!1,\half'}^e X''$ \\ 
\cline{5-8} 
\multicolumn{4}{|c||}{}   & $\Delta V=0$ & $\ket{\half,+\half}$ & odd & 
$\frac{1}{2\sqrt{2}} V_{\!0,\half}^o Y'' \sin\theta''$ \\ 
\cline{5-8} 
\multicolumn{4}{|c||}{}   & $\Delta V=0$ & $\ket{\half',+\half}$ & even & 
$-\frac{\sqrt{3}}{2} V_{\!0,\half'}^{\prime e} X''$ \\ 
\hline 
\end{tabular}
\caption{Comparison of decays of $\Bz_d$ and $\Bzb_s$ to the final state $\Kp\piz\pim$.}
\label{tab:KpPizPim}
\end{table*}

Let us now consider the decay $\Bz_d \text{ or } \Bzb_s \to \Kp\piz\pim$ in
which isospin symmetry allows the exchange of $\piz$ and $\pim$, and $V$-spin
symmetry allows exchange of $\Kp$ and $\piz$. This leads to the following
exchanges in the Dalitz plot:
\begin{align*}
V\text{-spin} \equiv \Kp \leftrightarrow \piz \implies s \leftrightarrow t,\\
\text{Isospin} \equiv \piz \leftrightarrow \pim \implies t \leftrightarrow u.
\end{align*}
Under exact isospin and $V$-spin, the final state $\Kp\piz\pim$ has,
the following two possibilities:
\begin{enumerate}
\item $\Kp\piz$ would exist in either symmetrical or anti-symmetrical
state w.r.t.\ their exchange in space, and
\item $\piz\pim$ would exist in either symmetrical or anti-symmetrical
state w.r.t.\ their exchange in space.
\end{enumerate}
Following the steps as enunciated in subsection~\ref{subsec:IU}, the amplitude
for the decay can be shown to have two components, one which is fully symmetric
under exchange of any pair of final particles, and the other fully
anti-symmetric under the same exchange.

The final state can be expanded in terms of isospin and $V$-spin states as
follows:
\begin{itemize}
\item Isospin
\begin{align}
\ket{\Kp\piz\pim} &= \frac{1}{\sqrt{5}} \ket{\frac{5}{2},-\half}_I^e +
\sqrt{\frac{3}{10}} \ket{\frac{3}{2}, - \half}_I^e \nn\\ %
& \quad + \frac{1}{\sqrt{6}} \ket{\frac{3}{2}, - \half}_I^o + \frac{1}{\sqrt{3}}
\ket{\half,-\half}_I^o,
\end{align}
where the superscripts $e,o$ denote even, odd behavior of the state under the
exchange $\piz \leftrightarrow \pim$. 

\item $V$-spin
\begin{align*}
\ket{\Kp\piz\pim} &= \frac{1}{2\sqrt{5}} \ket{\frac{5}{2},+\half}_V^e +
\frac{\sqrt{3}}{2\sqrt{10}} \ket{\frac{3}{2},+\half}_V^e \nn\\ %
& \quad + \frac{1}{2\sqrt{6}} \ket{\frac{3}{2},+\half}_V^o + \frac{1}{2\sqrt{3}}
\ket{\half,+\half}_V^o \nn\\ %
& \quad - \frac{1}{2} \ket{\frac{3'}{2},+\half}_V^e - \frac{1}{\sqrt{2}}
\ket{\frac{1'}{2},+\half}_V^e,
\end{align*}
where the superscripts $e,o$ denote even, odd behavior of the state under the
exchange $\Kp \leftrightarrow \piz$. 
\end{itemize}
The sign changes as can be noticed in the above states arise from exchange of
particles in the two particle states given below (as also noted in
Table~\ref{tab:modes}):
\begin{itemize}
\item Isospin:
\begin{align}
\ket{\piz\pim} &= \frac{1}{\sqrt{2}} \Big( \ket{2,-1}_I + \ket{1,-1}_I \Big),\\
\ket{\pim\piz} &= \frac{1}{\sqrt{2}} \Big( \ket{2,-1}_I - \ket{1,-1}_I \Big).
\end{align}
\item $V$-spin:
\begin{align}
\ket{\Kp\piz} &= -\frac{1}{2\sqrt{2}} \Big(\ket{2,+1}_V + \ket{1,+1}_V
\Big) + \frac{\sqrt{3}}{2}\ket{1',+1}_V,\\ %
\ket{\piz\Kp} &= -\frac{1}{2\sqrt{2}} \Big(\ket{2,+1}_V - \ket{1,+1}_V
\Big) + \frac{\sqrt{3}}{2}\ket{1',+1}_V.
\end{align}
\end{itemize}
It would be clear from the expressions above that if isospin were an exact
symmetry, the $\ket{2,-1}_I$ and $\ket{1,-1}_I$ states of $\ket{\pim\piz}$ would
exist in even and odd partial wave states respectively, as was the case in
subsection~\ref{subsec:IU} also. On the other hand, if $V$-spin were an exact
symmetry the state $\ket{\Kp\piz}$ must remain unchanged under $\Kp
\leftrightarrow \piz$ exchange. Under $V$-spin the $\ket{\Kp\piz}$ state can
exist in $\ket{2,+1}_V$ and $\ket{1,+1}_V$, out of which $\ket{1,+1}_V$ has a
contribution from the $\ket{0,0}_{V,8}$ admixture in $\piz$, denoted above by
$\ket{1',+1}_V$. Both state $\ket{2,+1}_V$ and the state $\ket{1',+1}_V$ exist
in space symmetric (even partial wave) states, and that part of $\ket{1,+1}_V$
arising out of $\ket{1,0}_V$ part of $\piz$ exists in space anti-symmetric (odd
partial wave) state.

If we consider the initial state to be $\Bz_d$ which is isospin
$\ket{\half,+\half}_I$ state but $V$-spin singlet $\ket{0,0}_V$ state, the
standard model Hamiltonian allows only $\Delta I = 0,1$ and $\Delta V = \half,
\thalf$ transitions. Therefore, in addition to the isospin amplitudes of
Eq.~\ref{eq:T}, we can define the following $V$-spin amplitudes:
\begin{align}
V_{\thalf,\thalf} &= \biggbracket{\thalf,\pm\frac{1}{2}}{\mathscr{ H}_{\Delta
V=\thalf}}{0,0}~, \\ %
V'_{\thalf,\thalf} &=\biggbracket{\thalf',\pm\frac{1}{2}}{\mathscr{ H}_{\Delta
V=\thalf}}{0,0}~, \\ %
\v&=\pm\sqrt{\frac{2}{3}}\biggbracket{\frac{1}{2},\pm\frac{1}{2}}{\mathscr{
H}_{\Delta V=\tfrac{1}{2}}}{0,0}~, \\ %
\vp{} &=\sqrt{\frac{1}{3}}\biggbracket{\frac{1'}{2},\pm\frac{1}{2}}{\mathscr{
H}_{\Delta V=\tfrac{1}{2}}}{0,0}.
\end{align}
The amplitude for the process $\Bz_d \to \Kp\piz\pim$ can, therefore, be
written as
\begin{align}
A(\Bz_d \to \Kp\piz\pim) &= -\frac{3}{\sqrt{10}} \T{3}^e X + \frac{1}{\sqrt{2}} 
\left( - \T{1}^o + \t^o \right) Y \sin\theta,\\
A(\Bz_d \to \Kp\piz\pim) &= \sqrt{\frac{3}{2}} \left( \frac{1}{\sqrt{20}} 
V_{\thalf,\thalf}^e - \frac{1}{\sqrt{6}} V_{\thalf,\thalf'}^{\prime e} - 
V_{\half, \half'}^{\prime e} \right) X'' \nn\\
& \quad + \frac{1}{2\sqrt{2}} \left( \frac{1}{\sqrt{3}} V_{\thalf,\thalf}^o + 
V_{\half,\half}^o \right) Y'' \sin\theta'',
\end{align}
where $X''$ and $Y''$ are functions that are, in general, arbitrary functions of
$r$ and $\cos\theta''$, and are introduced to take care of spatial and kinematic
contributions to the decay amplitude. As argued before, the part of the
amplitude even under isospin must also be even under $V$-spin and the part odd
under isospin must again be odd under $V$-spin:
\begin{align}
\label{eq:amp2-even} %
\ASS &= \frac{3}{\sqrt{10}} \T{3}^{\e}\,X \nn\\ %
&= \sqrt{\frac{3}{2}} \left( \frac{1}{\sqrt{20}} V_{\thalf,\thalf}^e -
\frac{1}{\sqrt{6}} V_{\thalf,\thalf'}^{\prime e} - V_{\half, \half'}^{\prime e}
\right) X'',\\ %
\label{eq:amp2-odd} %
\AAA &=\frac{1}{\sqrt{2}} (-\T{1}^{\o}+ \t^{\o} )\,Y\sin\theta \nn\\ %
&=\frac{1}{2\sqrt{2}} \left( \frac{1}{\sqrt{3}} V_{\thalf,\thalf}^o +
V_{\half,\half}^o \right) Y'' \sin\theta''.
\end{align}

We can conclude
that the Dalitz plot distribution in the even numbered sextants would be
identical to one another, and those of odd numbered sextants would also be
similar. Any deviation from this would constitute a signature of simultaneous
violations of isospin and $V$-spin.

Since $\Kp$ and $\pim$ belong to different multiplets of $U$-spin, in order to
consider the symmetry properties under $\Kp\leftrightarrow \pim$ one needs to
define the $G$-parity analogue of $U$-spin, denoted by $\GU$ and defined in the
Appendix~\ref{sec:appendix}. Since charge conjugation is a good symmetry in strong interaction,
$\GU$ is as good as $U$-spin itself. The state $\ket{\Kp\pim}$ is composed of
states which are even and odd under $\GU$-parity:
\begin{equation*}
\ket{\Kp\pim} = \frac{1}{2} \left( \ket{\Kp\pim}_\e + \ket{\Kp\pim}_\o
\right),
\end{equation*}
where 
\begin{align*}
\ket{\Kp\pim}_e &= \ket{\Kp\pim} - \ket{\pim\Kp},\\
\ket{\Kp\pim}_o &= \ket{\Kp\pim} + \ket{\pim\Kp},
\end{align*}
and
\begin{align*}
\GU \ket{\Kp\pim}_e &= \ket{\Kp\pim}_e,\\
\GU \ket{\Kp\pim}_o &= - \ket{\Kp\pim}_o.
\end{align*}
We have already proven that the amplitudes for the decay $\Bz_d\to\Kp\piz\pim$
has two parts one even and the other odd under the exchange of any two particles
in the final state. Hence, $\ASS$ is odd under $\GU$ and $\AAA$ is even under
$\GU$. Since the two $\GU$-parity amplitudes do not interfere the two amplitudes
$\ASS$ and $\AAA$ do not interfere in the Dalitz plot distribution resulting in
$f_{AA}$ being zero (Eq.~\eqref{eq:faa}). Therefore if $\GU$ is a good symmetry
of nature  it is interesting to conclude that the Dalitz plot is completely
symmetric under $s\leftrightarrow t \leftrightarrow u$. The Dalitz plot
asymmetries which would be a measure of the extent of breaking of the $\SU(3)$
flavor symmetry are, therefore, given by
\begin{align}
\mathbb{A}_{\text{Isospin}} &= \modulus{\frac{\Sigma^{I}_{VI} -
\Sigma^{III}_{IV}}{\Sigma^{I}_{VI} + \Sigma^{III}_{IV}}} +
\modulus{\frac{\Sigma^{III}_{IV} - \Sigma^{V}_{II}}{\Sigma^{III}_{IV} +
\Sigma^{V}_{II}}} + \modulus{\frac{\Sigma^{V}_{II} -
\Sigma^{I}_{VI}}{\Sigma^{V}_{II} + \Sigma^{I}_{VI}}} \nn\\ %
& \quad + \modulus{\frac{\Delta^{I}_{VI} - \Delta^{III}_{IV}}{\Delta^{I}_{VI} +
\Delta^{III}_{IV}}} + \modulus{\frac{\Delta^{III}_{IV} -
\Delta^{V}_{II}}{\Delta^{III}_{IV} + \Delta^{V}_{II}}} +
\modulus{\frac{\Delta^{V}_{II} - \Delta^{I}_{VI}}{\Delta^{V}_{II} +
\Delta^{I}_{VI}}},\\
\mathbb{A}_{U\text{-spin}} &= \modulus{\frac{\Sigma^{V}_{IV} -
\Sigma^{I}_{II}}{\Sigma^{V}_{IV} + \Sigma^{I}_{II}}} +
\modulus{\frac{\Sigma^{I}_{II} - \Sigma^{III}_{VI}}{\Sigma^{I}_{II} +
\Sigma^{III}_{VI}}} + \modulus{\frac{\Sigma^{III}_{VI} -
\Sigma^{V}_{IV}}{\Sigma^{III}_{VI} + \Sigma^{V}_{IV}}} \nn\\ %
& \quad + \modulus{\frac{\Delta^{V}_{IV} - \Delta^{I}_{II}}{\Delta^{V}_{IV} +
\Delta^{I}_{II}}} + \modulus{\frac{\Delta^{I}_{II} -
\Delta^{III}_{VI}}{\Delta^{I}_{II} + \Delta^{III}_{VI}}} +
\modulus{\frac{\Delta^{III}_{VI} - \Delta^{V}_{IV}}{\Delta^{III}_{VI} +
\Delta^{V}_{IV}}},\\
\mathbb{A}_{V\text{-spin}} &= \modulus{\frac{\Sigma^{I}_{IV} -
\Sigma^{III}_{II}}{\Sigma^{I}_{IV} + \Sigma^{III}_{II}}} +
\modulus{\frac{\Sigma^{III}_{II} - \Sigma^{V}_{VI}}{\Sigma^{III}_{II} +
\Sigma^{V}_{VI}}} + \modulus{\frac{\Sigma^{V}_{VI} -
\Sigma^{I}_{IV}}{\Sigma^{V}_{VI} + \Sigma^{I}_{IV}}} \nn\\ %
& \quad + \modulus{\frac{\Delta^{I}_{IV} - \Delta^{III}_{II}}{\Delta^{I}_{IV} +
\Delta^{III}_{II}}} + \modulus{\frac{\Delta^{III}_{II} -
\Delta^{V}_{VI}}{\Delta^{III}_{II} + \Delta^{V}_{VI}}} +
\modulus{\frac{\Delta^{V}_{VI} - \Delta^{I}_{IV}}{\Delta^{V}_{VI} +
\Delta^{I}_{IV}}},
\end{align}
where the $\Sigma$'s and $\Delta$'s are as defined in Eqs.~\eqref{eq:Sigma} and
\eqref{eq:Delta} respectively. It must again be noted that these asymmetries are
in general functions of $r$ and $\theta$ (or $\theta'$ or $\theta''$), and are
defined throughout the Dalitz plot region, including resonant regions. It would
again be interesting to look for patterns in the variations of these asymmetries
inside the Dalitz plot. Observation of these asymmetries would quantify the
extent of breaking of $\SU(3)$  flavor symmetry in the concerned decay mode. One
can also look for such asymmetries in the Dalitz plot for $\Bzb_s \to
\Kp\piz\pim$. The amplitudes for this process are given in Table~\ref{tab:KpPizPim}.

\subsection{Decay Mode with final state \texorpdfstring{$\Kp\piz\Kzb$}{K+ pi0
K0bar}}\label{subsec:UV}

\begin{table*}
\centering
\begin{tabular}{|c|c|c|l||c|c|c|l|}
\hline 
\multicolumn{8}{|c|}{$\Bp \to \Kp\piz\Kzb$} \\ \hline
\multicolumn{4}{|c||}{$U$-spin \Big(initial state $\ket{0,0}$\Big)} & 
\multicolumn{4}{c|}{$V$-spin \Big(initial state $\ket{\half,+\half}$\Big)} \\ 
\hline 
transition & final state & symmetry & Amplitude & transition & final state & 
symmetry & Amplitude \\ 
\hline 
 $\Delta U =\half$ & $\ket{\half,-\half}$ & odd & $\frac{1}{2\sqrt{2}} 
 U_{\!\half,\half}^o Y' \sin\theta'$ &  $\Delta V=1$ & $\ket{\thalf,+\half}$ & 
 mixed & $-\frac{3}{2\sqrt{10}} V_{\!1,\thalf}^e X'' - \frac{1}{2\sqrt{2}} 
 V_{\!1,\thalf}^o Y'' \sin\theta''$ \\ 
\hline
 $\Delta U =\half$ & $\ket{\half', -\half}$ & even & $\frac{\sqrt{3}}{\sqrt{2}} 
 U_{\!\half,\half'}^{\prime e} X'$   & $\Delta V=1$ & $\ket{\thalf',+\half}$ & 
 even & $\frac{\sqrt{3}}{2}V_{\!1,\thalf'}^{\prime e} X''$ \\ 
\hline
$\Delta U =\thalf$ & $\ket{\thalf,-\half}$ & mixed & 
$\frac{\sqrt{3}}{2\sqrt{10}} U_{\!\thalf,\thalf}^e X' - \frac{1}{2\sqrt{6}} 
U_{\!\thalf,\thalf}^o Y' \sin\theta'$ & $\Delta V= 1$ & $\ket{\half,+\half}$ & 
odd & $-\frac{1}{2\sqrt{2}} V_{\!1,\half}^o Y'' \sin\theta''$ \\ 
\hline
  $\Delta U =\thalf$ & $\ket{\thalf',-\half}$ & even & $\half 
  U_{\!\half,\half'}^{\prime e} X'$ &  $\Delta V= 1$ & $\ket{\half',+\half}$ & 
  even & $\frac{\sqrt{3}}{2} V_{\!1,\half'}^{\prime e} X''$ \\ 
\hline
\multicolumn{4}{|c||}{} & $\Delta V =0$ & $\ket{\half,+\half}$ & odd & 
$-\frac{1}{2\sqrt{2}} V_{0,\half}^o Y'' \sin\theta''$\\
\cline{5-8}
\multicolumn{4}{|c||}{}  & $\Delta V =0$ & $\ket{\half',+\half}$ & even & 
$\frac{\sqrt{3}}{2} V_{0,\half}^{\prime e} X''$\\
\hline \hline
\multicolumn{8}{|c|}{$\Dp \to \Kp\piz\Kzb$} \\ \hline
\multicolumn{4}{|c||}{$U$-spin \Big(initial state $-\ket{\half,-\half}$\Big)} & 
\multicolumn{4}{c|}{$V$-spin \Big(initial state $\ket{0,0}$\Big)} \\ 
\hline 
transition & final state & symmetry & Amplitude &  transition & final state & 
symmetry & Amplitude \\ 
\hline 
$\Delta U =1$ & $\ket{\thalf,-\half}$ & mixed & 
$-\frac{3}{2\sqrt{10}} U_{\!1,\thalf}^e X' -\frac{1}{2\sqrt{2}} 
U_{\!1,\thalf}^o Y' \sin\theta'$ & $\Delta 
V=\thalf$ & $\ket{\thalf,+\half}$ & mixed & $-\frac{\sqrt{3}}{2\sqrt{10}} 
V_{\!\thalf,\thalf}^e X'' - \frac{1}{2\sqrt{6}} V_{\!\thalf,\thalf}^o Y'' 
\sin\theta''$ \\ 
\hline
  $\Delta U =1$ & $\ket{\thalf', -\half}$ & even & $\frac{\sqrt{3}}{2} 
  U_{\!1,\thalf'}^{\prime e} X'$ & $\Delta V=\thalf$ & $\ket{\thalf',+\half}$ & 
  even & $\half V_{\!\thalf,\thalf'}^{\prime e} X''$ \\ 
\hline
 $\Delta U =1$ & $\ket{\half,-\half}$ & odd & $\frac{1}{2\sqrt{2}} 
 U_{\!1,\half}^o Y' \sin\theta'$ & $\Delta V=\half$ & $\ket{\half,+\half}$ & 
 odd & $-\half V_{\!\half,\half}^o Y'' \sin\theta''$ \\ 
\hline
 $\Delta U = 1$ & $\ket{\half',-\half}$ & even & $-\frac{\sqrt{3}}{2} 
 U_{1,\half'}^{\prime e} X'$ & $\Delta V=\half$ & $\ket{\half',+\half}$ & even 
 & $\frac{\sqrt{3}}{\sqrt{2}} V_{\!\half,\half'}^{\prime e} X''$ \\ 
\hline
$\Delta U =0$ & $\ket{\half,-\half}$ & odd & $-\frac{1}{2\sqrt{2}} 
U_{\!0,\half}^o Y' \sin\theta'$ &  \multicolumn{4}{c|}{} \\ 
\cline{1-4}
 $\Delta U = 0$ & $\ket{\half',-\half}$ & even & $\frac{\sqrt{3}}{2} 
 U_{0,\half'}^{\prime e} X'$ &  \multicolumn{4}{c|}{} \\ 
\hline 
\end{tabular}
\caption{Comparison of amplitudes for the decays of $\Bp$ and $\Dp$ to the final state $\Kp\piz\Kzb$.}
\label{tab:KpPizKzb}
\end{table*}

For study of simultaneous violations of both $U$-spin
and $V$-spin, we look at decays such as $\Bp \text{ or } \Dp \to \Kp \piz\Kzb$
and their conjugate modes. In this state, $\Kp$ and $\piz$ are exchangeable
under $V$-spin and $\piz,\Kzb$ are exchangeable under $U$-spin. Under $V$-spin,
the $\Kp\piz$ state can exist in $\ket{2,+1}_V$ and $\ket{1,+1}_V$, out of which
the state $\ket{1,+1}_V$ has a contribution from the $\ket{0,0}_{V,8}$ admixture
in $\piz$. Thus assuming $V$-spin to be an exact symmetry would put the state
$\ket{2,+1}_V$ and that part of $\ket{1,+1}_V$ state coming from $\ket{0,0}_{V,8}$
contribution of $\piz$ in space symmetric (even partial wave) state. The
remaining part of $\ket{1,+1}_V$ state would be in space anti-symmetric (odd
partial wave) state. Similarly, the $\piz\Kzb$ state would exist in
$\ket{2,-1}_U$ and $\ket{1,-1}_U$, out of which the state $\ket{1,-1}_U$ has a
contribution from the $\ket{0,0}_{U,8}$ admixture in $\piz$. Thus, if $U$-spin
were assumed to be an exact symmetry, the states $\ket{2,-1}_U$ and the
$\ket{1,-1}_U$ state coming from $\ket{0,0}_{U,8}$ part of $\piz$ would exist in
space symmetric (even partial wave) states, and the other part of $\ket{1,-1}_U$
would exist in space anti-symmetric (odd partial wave) state.

Therefore, under exact $U$-spin and $V$-spin, the final state $\Kp\piz\Kzb$ has,
the following two possibilities:
\begin{enumerate}
\item $\Kp\piz$ would exist in either symmetrical or anti-symmetrical
state w.r.t.\ their exchange in space, and
\item $\piz\Kzb$ would exist in either symmetrical or anti-symmetrical
state w.r.t.\ their exchange in space.
\end{enumerate}
Again, following the steps as enunciated in subsection~\ref{subsec:IU} we can
conclude that the Dalitz plot distribution in the even numbered sextants would
be identical to one another, and those of odd numbered sextants would also be
similar, as given in Eqs.~\eqref{eq:fodd} and \eqref{eq:feven}. Any deviation
from this would constitute a signature of simultaneous violations of $U$-spin
and $V$-spin. We can once again reaffirm the same logic as given in
subsections~\ref{subsec:IU} and \ref{subsec:IV}, by invoking the $\GI$-parity
operator (see Appendix~\ref{sec:appendix}) to connect $\Kp$ and $\Kzb$ belonging
to two different isospin doublets. This would lead to a fully symmetric Dalitz
plot which would be broken when $\GI$ is broken. The amplitudes for the two
decay modes under consideration are given in Table~\ref{tab:KpPizKzb}. The
Dalitz plot asymmetries that can be useful in this case are given by
\begin{align}
\mathbb{A}_{\text{Isospin}} &= \modulus{\frac{\Sigma^{V}_{IV} -
\Sigma^{I}_{II}}{\Sigma^{V}_{IV} + \Sigma^{I}_{II}}} +
\modulus{\frac{\Sigma^{I}_{II} - \Sigma^{III}_{VI}}{\Sigma^{I}_{II} +
\Sigma^{III}_{VI}}} + \modulus{\frac{\Sigma^{III}_{VI} -
\Sigma^{V}_{IV}}{\Sigma^{III}_{VI} + \Sigma^{V}_{IV}}} \nn\\ %
& \quad + \modulus{\frac{\Delta^{V}_{IV} - \Delta^{I}_{II}}{\Delta^{V}_{IV} +
\Delta^{I}_{II}}} + \modulus{\frac{\Delta^{I}_{II} -
\Delta^{III}_{VI}}{\Delta^{I}_{II} + \Delta^{III}_{VI}}} +
\modulus{\frac{\Delta^{III}_{VI} - \Delta^{V}_{IV}}{\Delta^{III}_{VI} +
\Delta^{V}_{IV}}},\\
\mathbb{A}_{U\text{-spin}} &= \modulus{\frac{\Sigma^{I}_{VI} -
\Sigma^{III}_{IV}}{\Sigma^{I}_{VI} + \Sigma^{III}_{IV}}} +
\modulus{\frac{\Sigma^{III}_{IV} - \Sigma^{V}_{II}}{\Sigma^{III}_{IV} +
\Sigma^{V}_{II}}} + \modulus{\frac{\Sigma^{V}_{II} -
\Sigma^{I}_{VI}}{\Sigma^{V}_{II} + \Sigma^{I}_{VI}}} \nn\\ %
& \quad + \modulus{\frac{\Delta^{I}_{VI} - \Delta^{III}_{IV}}{\Delta^{I}_{VI} +
\Delta^{III}_{IV}}} + \modulus{\frac{\Delta^{III}_{IV} -
\Delta^{V}_{II}}{\Delta^{III}_{IV} + \Delta^{V}_{II}}} +
\modulus{\frac{\Delta^{V}_{II} - \Delta^{I}_{VI}}{\Delta^{V}_{II} +
\Delta^{I}_{VI}}},\\
\mathbb{A}_{V\text{-spin}} &= \modulus{\frac{\Sigma^{I}_{IV} -
\Sigma^{III}_{II}}{\Sigma^{I}_{IV} + \Sigma^{III}_{II}}} +
\modulus{\frac{\Sigma^{III}_{II} - \Sigma^{V}_{VI}}{\Sigma^{III}_{II} +
\Sigma^{V}_{VI}}} + \modulus{\frac{\Sigma^{V}_{VI} -
\Sigma^{I}_{IV}}{\Sigma^{V}_{VI} + \Sigma^{I}_{IV}}} \nn\\ %
& \quad + \modulus{\frac{\Delta^{I}_{IV} - \Delta^{III}_{II}}{\Delta^{I}_{IV} +
\Delta^{III}_{II}}} + \modulus{\frac{\Delta^{III}_{II} -
\Delta^{V}_{VI}}{\Delta^{III}_{II} + \Delta^{V}_{VI}}} +
\modulus{\frac{\Delta^{V}_{VI} - \Delta^{I}_{IV}}{\Delta^{V}_{VI} +
\Delta^{I}_{IV}}}.
\end{align}
Once again the asymmetries being, in general, functions of $r$ and $\theta$ (or
$\theta'$ or $\theta''$) it would be quite interesting to look for their
variation across the Dalitz plot. These would be the visible signatures of the
breaking of $\SU(3)$ flavor symmetry.

\subsection{Decay Mode with final state \texorpdfstring{$\pip\piz\Kzb$}{pi+ pi0
K0bar}}\label{subsec:IUV}

Finally, we consider a mode where each pair of particles in the final states can
be directly related by one of the three $\SU(2)$ symmetries, namely isospin,
$U$-spin and $V$-spin. Here we do not need $\GI$, $\GV$ or $\GU$ to relate the
final states. We consider as an example decays with final state $\pip\piz\Kzb$
such as $\Dp \to \pip\piz\Kzb$ and the conjugate mode. In the final state
considered here, isospin exchange implies $\piz \leftrightarrow \pip$, $U$-spin
exchange implies $\piz \leftrightarrow \Kzb$ and $V$-spin exchange implies $\pip
\leftrightarrow \Kzb$. The $\SU(2)$ decompositions of all the pairs of particles
under their respective $\SU(2)$ symmetries have already been considered in
subsections~\ref{subsec:IU}, \ref{subsec:IV}, \ref{subsec:UV}. Once again, the
steps elaborated in subsection~\ref{subsec:IU} are applicable to this case also.
The amplitudes for this decay mode can be easily read off from
Table~\ref{tab:PipPizKzb}. However, in this mode the  even and odd contributions
to the decay amplitude can interfere as they are not eigenstates of $\GV$,
resulting in even and odd numbered sextants to have distinctly different density
of events as depicted in Eqs.~\eqref{eq:fodd} and \eqref{eq:feven}.  Note that
the Dalitz plot distributions for the even (odd) numbered sextants of the Dalitz
plot would still be identical if isospin and $U$-spin are exact symmetries. The
breakdown of isospin, $U$-spin and  $V$-spin could be quantitatively measured
using the following asymmetries:
\begin{align}
\mathbb{A}_{\text{Isospin}} &= \modulus{\frac{\Sigma^{I}_{IV} -
\Sigma^{III}_{II}}{\Sigma^{I}_{IV} + \Sigma^{III}_{II}}} +
\modulus{\frac{\Sigma^{III}_{II} - \Sigma^{V}_{VI}}{\Sigma^{III}_{II} +
\Sigma^{V}_{VI}}} + \modulus{\frac{\Sigma^{V}_{VI} -
\Sigma^{I}_{IV}}{\Sigma^{V}_{VI} + \Sigma^{I}_{IV}}} \nn\\ %
& \quad + \modulus{\frac{\Delta^{I}_{IV} - \Delta^{III}_{II}}{\Delta^{I}_{IV} +
\Delta^{III}_{II}}} + \modulus{\frac{\Delta^{III}_{II} -
\Delta^{V}_{VI}}{\Delta^{III}_{II} + \Delta^{V}_{VI}}} +
\modulus{\frac{\Delta^{V}_{VI} - \Delta^{I}_{IV}}{\Delta^{V}_{VI} +
\Delta^{I}_{IV}}},\\
\mathbb{A}_{U\text{-spin}} &= \modulus{\frac{\Sigma^{I}_{VI} -
\Sigma^{III}_{IV}}{\Sigma^{I}_{VI} + \Sigma^{III}_{IV}}} +
\modulus{\frac{\Sigma^{III}_{IV} - \Sigma^{V}_{II}}{\Sigma^{III}_{IV} +
\Sigma^{V}_{II}}} + \modulus{\frac{\Sigma^{V}_{II} -
\Sigma^{I}_{VI}}{\Sigma^{V}_{II} + \Sigma^{I}_{VI}}} \nn\\ %
& \quad + \modulus{\frac{\Delta^{I}_{VI} - \Delta^{III}_{IV}}{\Delta^{I}_{VI} +
\Delta^{III}_{IV}}} + \modulus{\frac{\Delta^{III}_{IV} -
\Delta^{V}_{II}}{\Delta^{III}_{IV} + \Delta^{V}_{II}}} +
\modulus{\frac{\Delta^{V}_{II} - \Delta^{I}_{VI}}{\Delta^{V}_{II} +
\Delta^{I}_{VI}}},\\
\mathbb{A}_{V\text{-spin}} &= \modulus{\frac{\Sigma^{V}_{IV} -
\Sigma^{I}_{II}}{\Sigma^{V}_{IV} + \Sigma^{I}_{II}}} +
\modulus{\frac{\Sigma^{I}_{II} - \Sigma^{III}_{VI}}{\Sigma^{I}_{II} +
\Sigma^{III}_{VI}}} + \modulus{\frac{\Sigma^{III}_{VI} -
\Sigma^{V}_{IV}}{\Sigma^{III}_{VI} + \Sigma^{V}_{IV}}} \nn\\ %
& \quad + \modulus{\frac{\Delta^{V}_{IV} - \Delta^{I}_{II}}{\Delta^{V}_{IV} +
\Delta^{I}_{II}}} + \modulus{\frac{\Delta^{I}_{II} -
\Delta^{III}_{VI}}{\Delta^{I}_{II} + \Delta^{III}_{VI}}} +
\modulus{\frac{\Delta^{III}_{VI} - \Delta^{V}_{IV}}{\Delta^{III}_{VI} +
\Delta^{V}_{IV}}}.
\end{align}
Once again these asymmetries being, in general, functions of $r$ and $\theta$
(or $\theta'$ or $\theta''$) it would be very interesting to look for their
variation across the Dalitz plot. These would constitute the visible signatures
of the breaking of $\SU(3)$ flavor symmetry.

\begin{table*}
\centering
\begin{tabular}{|c|c|c|l||c|c|c|l|}
\hline 
\multicolumn{8}{|c|}{$\Dp \to \pip\piz\Kzb$} \\ \hline
\multicolumn{4}{|c||}{Isospin \Big(initial state $\ket{\half,+\half}$\Big)} & 
\multicolumn{4}{c|}{$U$-spin \Big(initial state $-\ket{\half,-\half}$\Big)}  \\ 
\hline 
 transition & final state & symmetry & Amplitude &  transition & final state & 
 symmetry & Amplitude \\ 
\hline 
\multirow{2}{*}{$\Delta I =1$} & \multirow{2}{*}{$\ket{\thalf,+\thalf}$} & \multirow{2}{*}{mixed} & 
\multirow{2}{*}{$\frac{\sqrt{3}}{\sqrt{10}}T_{\!1,\thalf}^e X + \frac{\sqrt{3}}{\sqrt{2}} T_{\!1,\thalf}^o Y 
 \sin\theta$} &  \multirow{2}{*}{$\Delta U=1$} & $\ket{\thalf,-\thalf}$ & mixed & 
 $\frac{\sqrt{3}}{2\sqrt{5}} U_{\!1,\thalf}^e X' - \frac{\sqrt{3}}{2} U_{\!1,\thalf}^o Y' \sin\theta'$ \\ 
\cline{6-8}
 & & & & & $\ket{\thalf',-\thalf}$ & even & 
 $\frac{3}{2} U_{\!1,\thalf'}^{\prime e} X'$ \\ 
\hline
\end{tabular}
\caption{Amplitudes for the decay $\Dp \to \pip\piz\Kzb$. The $V$-spin
amplitudes can be written in a similar manner. For brevity we have not written
them explicitly.} \label{tab:PipPizKzb}
\end{table*}

\section{Conclusion}
\label{sec:conclusions}

In this paper we have elucidated a new model independent method to look for the
breaking of the $\SU(3)$ flavor symmetry in many three-body decay modes, namely
$\Bp \text{ or } \Dp_s \to \Kz\piz\pip$, $\Bz_d \text{ or }\Bzb_s \to
\Kp\piz\pim$, $\Bp \text{ or } \Dp \to \Kp\piz\Kzb$ and $\Dp \to \pip\piz\Kzb$.
The novelty in choosing these decay modes is that pairs of the final state do
belong to at least two different $\SU(2)$ triplets, and hence under the
assumption of exact $\SU(3)$ flavor symmetry, the amplitude for the process has
two parts: one fully symmetric and another fully anti-symmetric under the
exchanges $s \leftrightarrow t \leftrightarrow u$. This gives rise to a
characteristic pattern in the Dalitz plot distribution: the alternate 
sextants must have identical distribution of events. Any
deviation from this behavior would constitute an evidence for the breaking of
$\SU(3)$ flavor symmetry, which indeed is broken in nature. We have
provided mode specific Dalitz plot asymmetries which can be used to quantify the
extent of $\SU(3)$ symmetry breaking in each of the decay modes under our
consideration. These asymmetries are defined in the full region of the
Dalitz plot and  can be measured both along resonances and in the
non-resonant regions. A quantitative estimate  of the variation of these
asymmetries obtained experimentally would provide a valuable understanding of
$\SU(3)$ breaking effects. It would also be interesting to find regions of the
Dalitz plots where $\SU(3)$ is a good symmetry.  A better understanding and
measured estimate of $\SU(3)$ breaking would help in reliably estimating
hadronic uncertainties and hence result in effectively using it to measure weak
phases and search for new physics effects beyond the standard model.

\acknowledgments
N. G. D. thanks The Institute of Mathematical Sciences, Chennai,  for 
hospitality, where 
part 
of the work was done.
\vskip 0.5cm

\appendix

\section{\texorpdfstring{$G$}{G}-parity and final states}
\label{sec:appendix}

The $G$-parity operator $\GI$ (or $\GU$ or $\GV$) is defined as a rotation
through $\pi$ radian ($180^{\circ}$) around the second axis of isospin (or
$U$-spin or $V$-spin) space, followed by  charge conjugation ($\mathcal{C}$):
$\GI = \mathcal{C} e^{i\pi I_2} = \mathcal{C} e^{i\pi \tau_2/2}, $
where $I_2$ is the second generator of $\SU(2)$ isospin (or $U$-spin or 
$V$-spin) group, and $\tau_2$ is
the second Pauli matrix.
$G$-parity as defined here transforms the
various $\SU(2)$ multiplets as follows:
\begin{align*}
& \GI
\begin{pmatrix}
\pip \\ \piz \\ \pim
\end{pmatrix}
= -
\begin{pmatrix}
\pip \\ \piz \\ \pim
\end{pmatrix},
&& \GI
\begin{pmatrix}
\Kp \\ \Kz
\end{pmatrix}
= 
\begin{pmatrix}
\Kzb \\ - \Km
\end{pmatrix},
&& \GI 
\begin{pmatrix}
\Kzb \\ - \Km
\end{pmatrix}
= -
\begin{pmatrix}
\Kp \\ \Kz
\end{pmatrix},\\
& \GU
\begin{pmatrix}
\Kz \\ \piz \\ \Kzb
\end{pmatrix}
= -
\begin{pmatrix}
\Kz \\ \piz \\ \Kzb
\end{pmatrix},
&& \GU
\begin{pmatrix}
\Kp \\ \pip
\end{pmatrix}
= 
\begin{pmatrix}
\pim \\ - \Km
\end{pmatrix},
&& \GU 
\begin{pmatrix}
\pim \\ - \Km
\end{pmatrix}
= -
\begin{pmatrix}
\Kp \\ \pip
\end{pmatrix},\\
& \GV
\begin{pmatrix}
\Kp \\ \piz \\ \Km
\end{pmatrix}
= -
\begin{pmatrix}
\Kp \\ \piz \\ \Km
\end{pmatrix},
&& \GV
\begin{pmatrix}
\pip \\ \Kzb
\end{pmatrix}
= 
\begin{pmatrix}
\Kz \\ - \pim
\end{pmatrix},
&& \GV 
\begin{pmatrix}
\Kz \\ - \pim
\end{pmatrix}
= -
\begin{pmatrix}
\pip \\ \Kzb
\end{pmatrix},
\end{align*}

\end{document}